# On Radio Detection of Ultra-High Energy Neutrinos in Antarctic Ice

George M. Frichter [1], John P. Ralston and Douglas W. McKay
Department of Physics and Astronomy, and
Kansas Institute for Theoretical and Computational Science,
University of Kansas,
Lawrence, KS 66045, USA

July 20, 1995

[1]After 9-1-95: Bartol Research Institute, University of Delaware, Newark, DE 19716, USA

## Abstract

Interactions of ultrahigh energy neutrinos of cosmological origin in large volumes of dense, radio-transparent media can be detected via coherent Cherenkov emission from accompanying electromagnetic showers. Antarctic ice meets the requirements for an efficient detection medium for a radio frequency neutrino telescope. We carefully estimate the sensitivity of realistic antennas embedded deep in the ice to 100 MHz - 1 GHz signals generated by predicted neutrino fluxes from active galactic nuclei. Our main conclusion is that a *single radio receiver* can probe a $\sim 1$ km$^3$ volume for events with primary energy near 2 PeV and that the total number of events registered would be roughly 200 to 400 year$^{-1}$ in our most conservative estimate. An array of such receivers would increase sensitivity dramatically. A radio neutrino telescope could directly observe and test our understanding of the most powerful particle accelerators in the universe, simultaneously testing the standard theory of particle physics at unprecedented energies.

# 1 Introduction

An ultrahigh energy (UHE) neutrino telescope is a multipurpose instrument by its very nature. It has the potential to probe, in a fundamentally new manner, the most distant and energetic objects in the universe (such as active galactic nuclei or AGN) as well as exotic structures within our own galaxy [1]. Due to the enormous energies of the particle interactions involved ($> 10^{12}$eV), it also serves as a natural laboratory to test our understanding of fundamental physics in the UHE regime.

As a new observational window on the universe, neutrino astronomy's ultimate impact on our understanding of astrophysical phenomena cannot be known in advance. The history of astronomical science is certainly encouraging: every new observational tool has produced unforeseen discoveries. In the case of neutrino astronomy, the prospects seem to be particularly promising since for the first time one would use a fundamentally different particle to view the cosmos. Unlike photons, which are easily stopped by small amounts of intervening matter, neutrinos interact very weakly, as if custom-made to give us information about regions of the universe which can not be directly probed using the techniques of conventional astronomy. Of course, this same property also makes neutrino observation a technical and scientific challenge [1, 2].

As an example of the particle physics aspect of a neutrino observatory, consider the fundamental neutrino-nucleon cross section, $\sigma^{\nu N}$. When $\sqrt{s}$, the invariant center of mass energy, is much larger than the W-boson mass, $M_W$, the magnitude of this cross section is directly determined by the number of quarks carrying a small fraction, $x$, of the total nucleon momentum in the center of momentum frame [3, 4, 5, 6]. The quark distribution comes from fundamental, non-perturbative, strong interaction processes. It follows that UHE neutrino detection probes not just the depths of the cosmos but also intricate details of hadronic structure and strong interaction physics. One can only speculate about what other interesting phenomena might present themselves at untested energy extremes. For example, it has been suggested [7] that tau neutrinos resulting from neutrino oscillations would produce a clear signal in a UHE detector, providing evidence for "new physics" if observed.

Several new high energy detectors utilizing optical sensor technology are now coming into operation. These operate mainly by detecting muons from charged-current muon-neutrino interactions. The muon's path must intersect a detection volume arrayed with photomultiplier tubes to amplify visible Cherenkov photons the muon emits. The long range of the muon is helpful in probing a large target volume. This detection strategy appears to be ideal for observing neutrinos in the GeV to TeV range and it is being employed in a variety of naturally occurring, dense and optically transparent environments such as deep freshwater lakes (NESTOR, Lake Baikal), deep ocean (DUMAND), and Antarctic ice (AMANDA) [1]. However, the combination of flux and detector efficiency severely limits this approach in the energy regime above 100 TeV.

Here we consider an attractive alternate detection strategy based on coherent radio Cherenkov emission from neutrino-induced electromagnetic showers [8, 9]. This approach was first developed and tested in the context of cosmic ray induced air showers [8, 9, 10]. The advantages of going to a dense radio-transparent medium are spectacular but have only been fully appreciated relatively recently. The radio power emitted coherently scales inversely with the radiation length of the medium *squared.* About 1 million times more radio energy is emitted by a shower in ice than in the old air shower experiments. Cold ice happens to be an abundant natural substance that is quite transparent to radio signals, having an attenuation length of 1 km or more at Antarctic temperatures. Thus, Markov and Zheleznykh [11] suggested using the Antarctic ice as a solid natural target for neutrino detection, and this idea has been subsequently developed by a number of authors [12, 13, 14, 15].

The radio power emitted coherently also goes like the primary shower energy squared [8, 9]. (This can be compared with emission from charged shower particles that add *incoherently* as in the case of visible light, which rises at most linearly with energy.) The power received at a distance $r$ from an event falls like $1/r^2$ from geometry (assuming attenuation can be neglected). As a rule of thumb, for radio the detection range increases proportionally to the primary energy. This is a very interesting effect. A range increasing linearly with energy translates into a target volume going like the *energy-cubed.* Earlier numerical estimates including realistic antennas and signal/noise = 1 found radio signals detectable at a distance as large as 1 kilometer if the energy were 1 PeV [12, 15]. By the scaling laws, one could also detect 100 TeV neutrinos as far away as 100 m, or 10 TeV neutrinos at 10 m. It follows that radio detection is not limited intrinsically to fantastically high energy thresholds. What radio does is offer a way to probe enormous target volumes if UHE neutrinos exist. For PeV neutrinos and above, radio can probe *cubic kilometer-scale volumes per detector element* in the near future at a comparatively low cost.

Considerable attention is being given now to "KM3" proposals for visible-light-based detection with target volumes of the order of $1\mathrm{km}^3$ [1, 16]. These proposals are ambitious and exciting. Due to the different scaling laws with energy the radio and optical approaches are complementary rather than in direct competition. Depending on the primary flux, we find that the energies at which radio detection is most exciting range from 100 TeV to 100 PeV, bridging a gap in conventional methods and considerably extending the range of energy. It is a potentially valuable coincidence that the medium chosen for the AMANDA optical array (pure $-50^o$C ice) is also beautifully transparent to radio signals up to the GHz range [17] where coherence is maintained. The benefits of integrating radio and optical detection seem to be very great indeed and should be carefully considered.

It might seem that a few radio antennas placed on the ice surface, either alone or in conjunction with a conventional array, would be a good way to extend existing studies to above the PeV range. Early calculations by Ralston and McKay [13] assumed this geometry and detection of upward-going showers ignoring attenuation. However, Earth-shadowing effects cut the flux of upward going neutrino events above 100 TeV severely [18]. A reliable estimate must include the angular dependence of the attenuation [15]. The flux of of "horizontal" neutrinos which penetrate with minimal attenuation actually dominates. We present here the first calculations in which the full effects of the geometry are taken into account and a variable antenna depth is included; we consider optimizing the detection by studying buried detectors with good horizontal acceptance as a function of depth, to take advantage of unattenuated sideways flux at 1 PeV and above. Buried detection has not been considered in detail before. The results represent a substantial improvement in the whole scheme.

It is natural to contemplate a single telescope facility employing both optical and radio detection to observe cosmic neutrinos over an enormous range of energies. However, radio detection has enough advantages to make an independent facility very attractive. We find several hundred events from electron neutrinos, per detector per year, using standard flux estimates. Backgrounds from atmospheric neutrinos are negligible. These topics are discussed in de-tail in the sections which follow.

Because of the unusual growth in detection efficiency with energy, we have focused on the energy region of 100 TeV and above. In this regime, much of the recent activity in neutrino astronomy has been stimulated by the prospect that AGN may be intense sources of energetic neutrinos [19, 20, 21, 22]. Theoretical results for "typical" AGN have been integrated to estimate the diffuse neutrino flux to be expected from all the distant active galaxies in the universe. In this work we estimate the sensitivity of radio detection in Antarctic ice to two different predictions for this isotropic or unresolved AGN flux. Our results show that even a single radio receiver is capable of detecting AGN at the current predicted flux levels.

Not only would AGN be detected, but we might also be able to distinguish between competing models of the underlying physics because of the different energy dependences for the predicted flux. Very generally, if a primary neutrino integral spectrum is falling as $E^{-\gamma}$ with an unknown spectral index $\gamma$, and the efficiency for detection $V(E)$ is a known, increasing function of energy, as it is in our case, then the distribution of events, $\frac{dN}{d\log E}$, will have a characteristic maximum at some energy $E_{\rm max}$. This permits a direct measurement of the unknown source spectrum. We will show that the unknown spectral index can be found from an observed value of $E_{\rm max}$ by evaluating $\gamma = \frac{d\log V}{d\log E} |_{E_{\rm max}}$. As the reader will see in subsequent sections, the two AGN models we consider have spectral indices that differ by roughly 1 [19, 20, 21]. Given our evaluation of the efficiency for radio detection, we predict an order of magnitude difference in the value of $E_{\rm max}$ between the two models. With even modest energy resolution, one could reasonably hope to distinguish between these two pictures of AGN physics.

Figure (1) is a sketch indicating the geometry of a typical detected neutrino event to help the reader follow the sequence of concepts and calculations developed in the various sections of this paper. Sections 2 and 3 present estimates for the UHE flux incident at the Earth's surface and a detailed account of how this nominal flux is modified by interactions as it passes through the Earth on its way to the region containing the detector. The sketch indicates a typical UHE neutrino that enters the detection region and has a charged-current interaction with a target nucleon; a calculation of the rate for these interactions is the topic of Section 4. The resulting charged final state lepton initiates an electromagnetic shower of charged particles and photons. The production rate of these cascades is the subject of Section 5. Electromag-netic cascade evolution produces a net excess of electrons compared with positrons that gives a coherent Cherenkov pulse at radio frequencies. This pulse is a ring-shaped structure, indicated in the sketch by dashes that propagates along the surface of a cone defined by the Cherenkov angle for the medium. The characteristics of the radio signal and a detailed treatment of its propagation from the site of the electromagnetic shower to a point where it intersects the antenna array is the topic of Section 6. In Sections 7 and 8 we present an analysis of how the radio pulse couples to a realistic antenna/receiver combination and how it is detected against a model for the noise background. Using a signal-to-noise criterion, we present in Section 9 a calculation of the effective detection volume per antenna that incorporates all the threshold effects relevant to detection. The calculations of shower rates and effective volumes come together in Section 10 where we show the detected number of events per year per antenna. In Section 11 we give a procedure for determining the source spectral index from experimental data. Finally Section 12 summarizes our results and estimates rates for a multi-unit neutrino telescope.

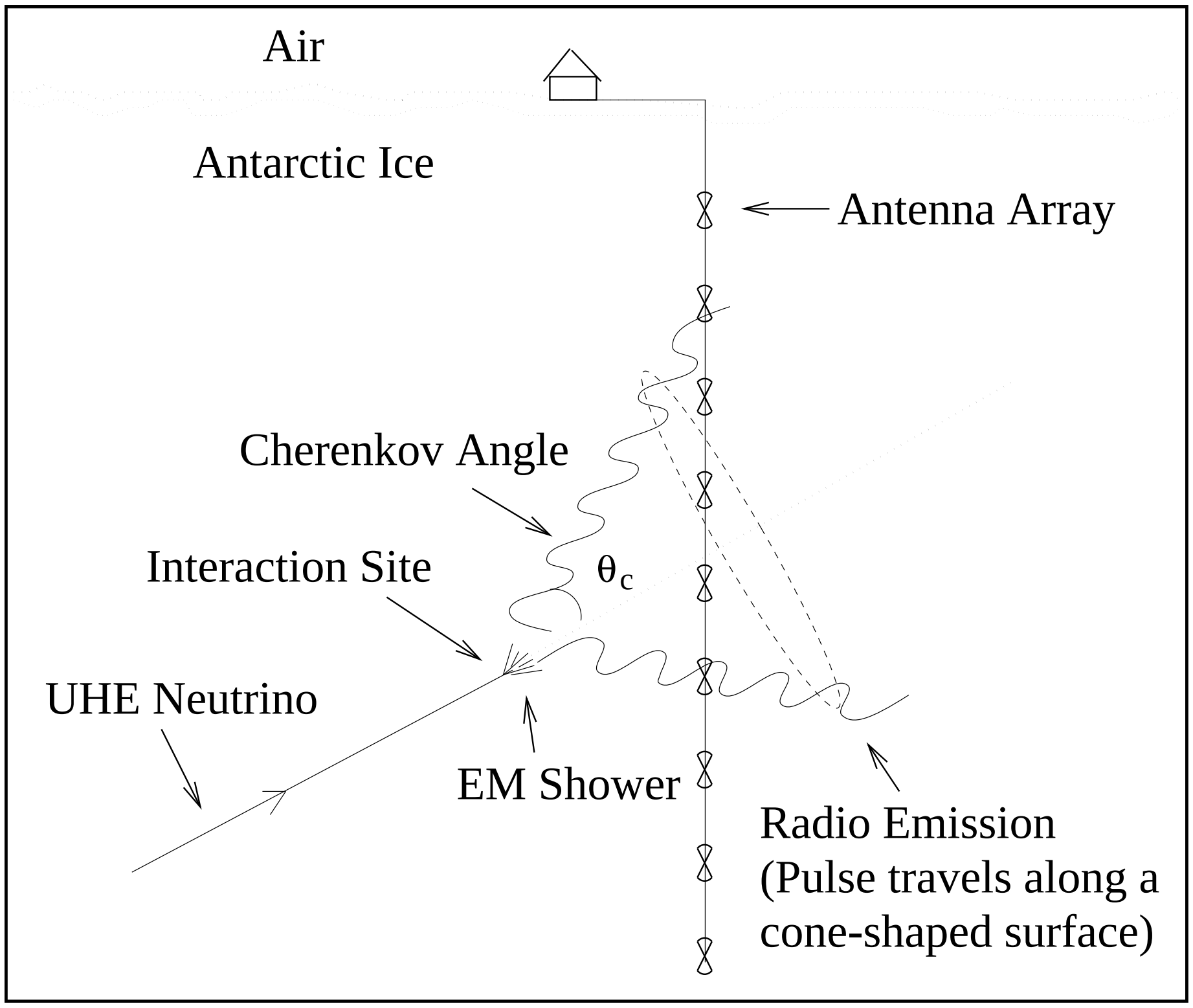

Figure 1: The geometry of a typical detected neutrino event. Our calculation includes all known effects: particle physics in the fundamental interaction cross sections, the angular dependence of attenuation in the Earth, electromagnetic shower evolution, coherent Cherenkov radio emission, polarization of the emitted field, pulse dispersion in the ice and antenna characteristics.

# 2 Predictions for Diffuse UHE Neutrino Fluxes

Figure (2) shows recent predictions for the unresolved AGN neutrino flux due to Szabo and Protheroe (denoted hereafter as SP) [20] and Stecker *et al.* (denoted ST)[19]. The Szabo and Protheroe flux we show is the most optimistic of their range of estimates. We have chosen these two AGN flux models as representative of the range of current theoretical prediction. Also shown is the atmospheric neutrino spectrum as predicted by Lipari (denoted AT)[23]. The the curves are piecewise linear approximations (on a log-log plot) to the published curves used for convenience. These three spectra will serve as the flux input for our calculations. The figure depicts the $\nu_\mu + \overline{\nu}_\mu$ flux at the Earth's surface. The ratio of electron neutrinos to muon neutrinos is approximately 2/3 in the SP model and 1/2 in the ST case.

We adopt the following notation. An isotropic differential neutrino flux at the Earth's surface can be expressed,

$$\phi(E_\nu) = \phi_o(E_o) * \frac{E_\nu}{E_o}^{-(\gamma(E_\nu)+1)} . \qquad (1)$$

Between the Earth's surface and the detector, these neutrinos must pass through different amounts of intervening matter. Cross sections for charged and neutral current interactions are sufficiently large at the energies of interest that we must ask how the originally isotropic flux appears after its journey.

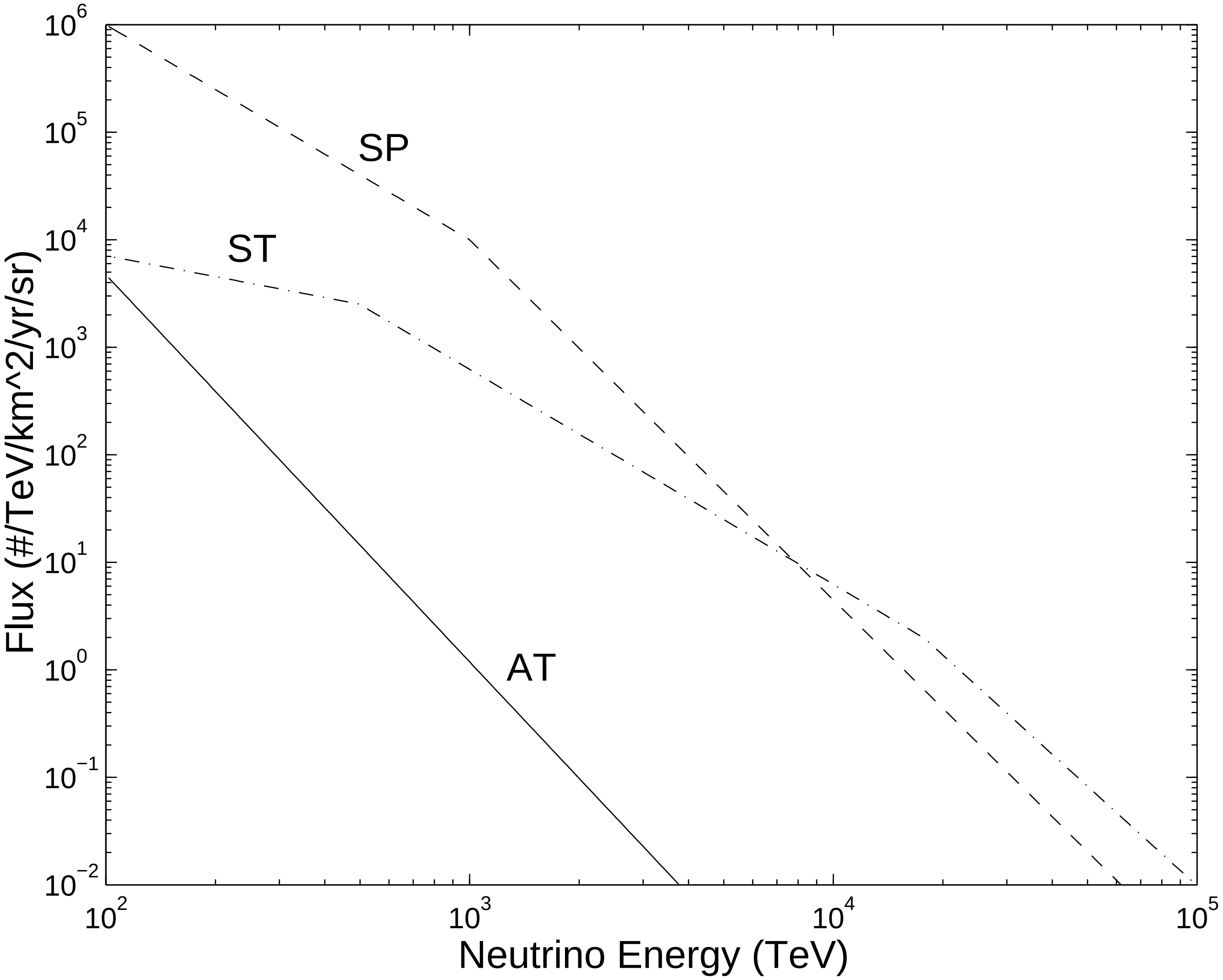

Figure 2: Three neutrino flux models used in the present calculation. Two spectra for diffuse AGN are due to Szabo and Protheroe (SP) and Stecker *et al.* (ST). The third curve is the atmospheric neutrino prediction of Lipari (AT).

## 3 Modification of Flux by Its Passage Through Earth

The neutrino flux which reaches a buried detector is modified by interactions with matter as it passes though the Earth. Aside from the reaction, $\overline{\nu}_e + e^- \rightarrow W^- \rightarrow$ anything, which is large only near the resonant energy 6.4 PeV, scattering of neutrinos on atomic electrons can safely be neglected, leaving charged and neutral current interactions with nucleons as the two relevant mechanisms to consider. The flux is modified in two distinct ways. Charged currents, $\nu + p \longrightarrow X + e^-$ and $\overline{\nu} + n \longrightarrow X + e^+$ directly remove neutrinos from the flux while neutral currents, $\nu + p \longrightarrow X + \nu$ and $\overline{\nu} + n \longrightarrow X + \overline{\nu}$ shift neutrinos into lower energy bins.

The effect is energy dependent owing to the energy dependence of the relevant interaction cross sections. We have recently calculated the charged current neutrino-nucleon cross section at ultra-high energy [6]. Our calculation is based on QCD evolution of recent electroproduction data at small Bjorken $x$ from the H1 and ZEUS collaborations at HERA [24, 25]. Figure (3) represents our current knowledge of the cross section over the GeV to PeV range showing previous estimates [3, 4, 5] for the UHE region that were based on incomplete knowledge of the small $x$ physics along with our improved calculation [6]. The region of existing high energy cross section data is indicated, along with one very high energy data point near 50 TeV extracted by the H1 group who measured $\sigma(ep \rightarrow \nu + X)$ [26]. The lone data point is at approximately 10 times the highest energy previously reported and is in good agreement with our new result, thus tying together two completely independent aspects of HERA measurements. Our result is roughly 2.2 times previous estimates for neutrino energies of 1 PeV and above. The following parameterization of our result is valid at the 5% level of accuracy over the range 50 TeV $< E <$ 50 PeV,

$$\sigma^{\nu N}(E_\nu) = (1.03\text{x}10^{-35}\text{cm}^2)(\frac{E_\nu}{\text{TeV}})^{0.822}\exp[-0.0231\,\ln^2(\frac{E_\nu}{\text{TeV}})]\,. \qquad (2)$$

At these large energies, the cross section is dominated by events in which the struck parton carries a very small fraction of the total nucleon momentum (*i.e.* small $x$). In this regime, the quark sea dominates over valence contributions and the equal abundance of quarks and anti-quarks means that the neutral and charged current cross section are proportional. The constant of proportionality is related to the weak mixing angle, a well known quantity.

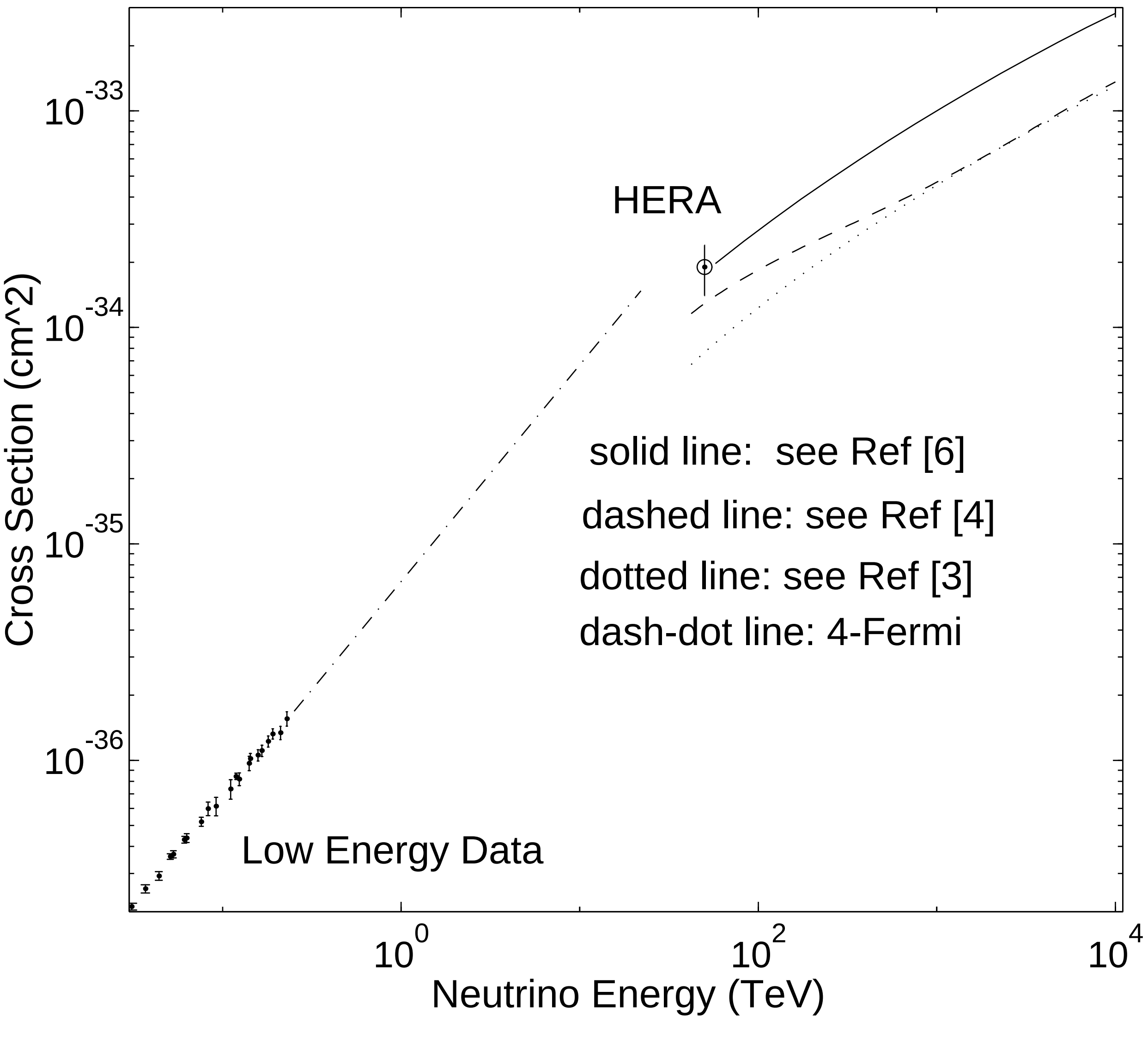

Figure 3: Total charged current neutrino-nucleon cross section versus incident neutrino energy. The extrapolated 4-Fermi result is shown along with the single high energy HERA data point. Also shown is the new ultrahigh energy prediction of Frichter, McKay and Ralston [ref. 6] along with previous results of McKay and Ralston [3] and Reno and Quigg [4].

$$\sigma_{\rm NC}^{\nu N}(E_\nu) \cong 0.30\ \sigma_{\rm CC}^{\nu N}(E_\nu) \tag{3}$$

In addition to the energy dependence, the modified flux gets an angular dependence due to differing amounts of matter it must traverse en route to the detector. The overall effect is expressed by an evolution equation for the flux. Consider charged and neutral current events occuring over an interval, $dt = n_N(z)dz$, with $n_N(z)$ being a local nucleon density for some position $z$ along the path,

$$\begin{aligned} \frac{d{\rm ln}\phi}{dt}(E_\nu) &= -\sigma_{\rm CC}^{\nu N}(E_\nu) - \sigma_{\rm NC}^{\nu N}(E_\nu) + \int_{E_\nu}^{\infty} dE'_\nu \frac{\phi(E'_\nu)}{\phi(E_\nu)} \frac{d\sigma_{\rm NC}^{\nu N}}{dE_\nu}(E'_\nu, E_\nu) \\ &\equiv -\sigma_{\rm eff}^{\nu N}(E_\nu, \gamma)\,. \end{aligned} \tag{4}$$

The effective neutrino-nucleon cross section, $\sigma_{\rm eff}^{\nu N}$, is a sum of charged and neutral current contributions evaluated at $E_\nu$, minus a correction which effectively reduces the cross section at the energy of interest because of neutral current interactions which occur at larger energies and have a final state neutrino energy $E_\nu$. Since the magnitude of the "$\nu$ regeneration" correction depends on the number of neutrinos available at energies larger than $E_\nu$, the effective cross section will be flux dependent via the spectral shape, $\gamma(E_\nu)$.

Now let $\theta$ be the nadir angle (the angle measured from downward vertical) of a particular primary neutrino momentum, and $n_N(z,\theta)$ be the nucleon density along the path as a function of $\theta$. The effects of the evolution can then be expressed simply in terms of the effective cross section [27],

$$\begin{aligned} \phi(E_\nu,\theta) &= \phi_o(E_\nu){\rm exp}[-\sigma_{\rm eff}(E_\nu,\gamma)t(\theta)] \\ &= \phi_o(E_\nu)k(E_\nu,\theta)\,, \end{aligned} \tag{5}$$

with $t(\theta) = \int n_N(z,\theta)dz$ and here $\phi_o(E_\nu)$ is the incident neutrino flux of Eq.(1). We make use of the PREM Earth structure model [28] to obtain $t(\theta)$.

In Fig.(4) we show effective cross sections for the SP and ST flux predictions along with the charged plus neutral current cross section to show the effect of the $\nu$ regeneration term of Eq.(4). The effect is largest for the ST flux model at energies below roughly 1 PeV owing to the spectrum's flatness compared with the SP flux. Higher energy neutrinos feed the lower energy flux more efficiently in the ST case than the SP case.

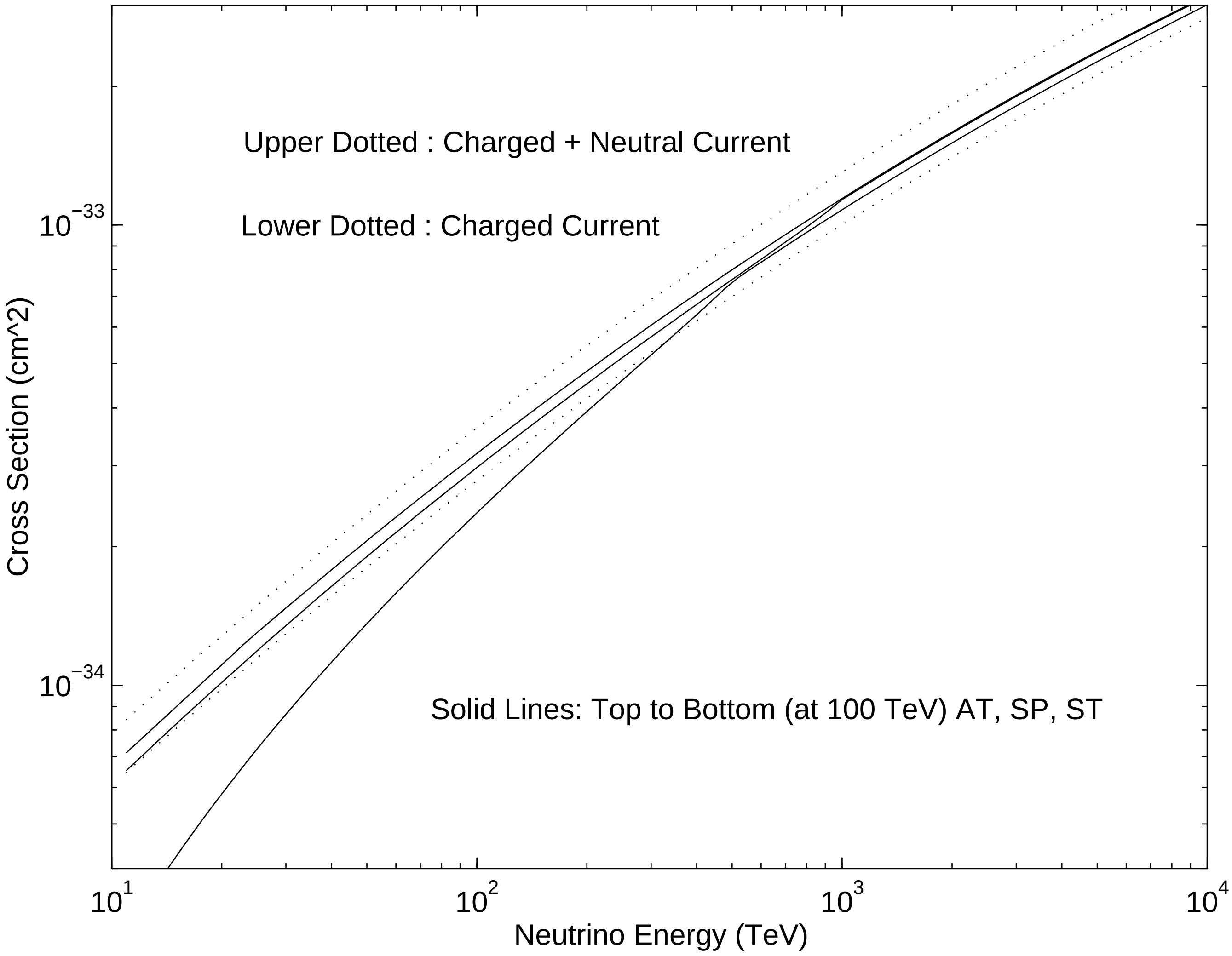

Figure 4: Effective cross section, $\sigma_{\rm eff}^{\nu N}$, for the SP and ST diffuse AGN flux models compared with $[\sigma_{\rm CC}^{\nu N} + \sigma_{\rm NC}^{\nu N}]$ to indicate the effect of neutrino regeneration.

With the effective cross sections in hand, we can use Eq.(5) to find the angular variation of the flux in the region of our detector. Figure (5) shows our result for the SP and ST AGN flux models. The angular variation of the attenuated flux clearly demonstrates the effect of shadowing by the Earth's mass which becomes ever more pronounced with increasing energy due to the rising interaction cross section of Eq.(2). In the calculations which follow, we assume that the flux as given by Eq.(5) is invariant with respect to translations within several kilometers of the detector. This is reasonable since such distances are very small compared with a typical mean free path for the neutrinos.

## 4 Charged-Current Lepton Production In Detection Region

Using the local neutrino flux, we can directly evaluate the number of leptons produced by charged-current interactions in the vicinity of the detector. We let $\phi^e$ and $\phi^\mu$ represent the local neutrino fluxes of electron and muon type as determined by Eq.(5). The rates, $\Gamma_{e,\mu}$, for electron and muon production in the neighborhood of the detector are found by integrating over all potential parent neutrinos having energy, $E_\nu$, weighted by the appropriate differential cross section for production of charged leptons with energy, $E_e$ or $E_\mu$:

$$\Gamma_{e,\mu}(E_{e,\mu},\theta) = \int_{E_{e,\mu}}^{\infty} dE_\nu \phi^{e,\mu}(E_\nu,\theta) n_N \frac{d\sigma_{\rm CC}^{\nu N}}{dE_{e,\mu}}(E_\nu, E_{e,\mu}) \,. \qquad (6)$$

Our results are shown in Fig.(6) and indicate the number of leptons produced per year, per energy bin, per unit of solid angle, per unit volume. Under the assumptions of the SP flux model, 40% of these are $e^\pm$ and the remaining 60% are $\mu^\pm$. In the ST case, the corresponding percentages are 33% and 66%. Each charged lepton has the potential to initiate an electromagnetic (EM) shower in the ice that is detectable by virtue of its accompanying Cherenkov shock wave. Our evaluation of the EM shower rate is the topic of the next

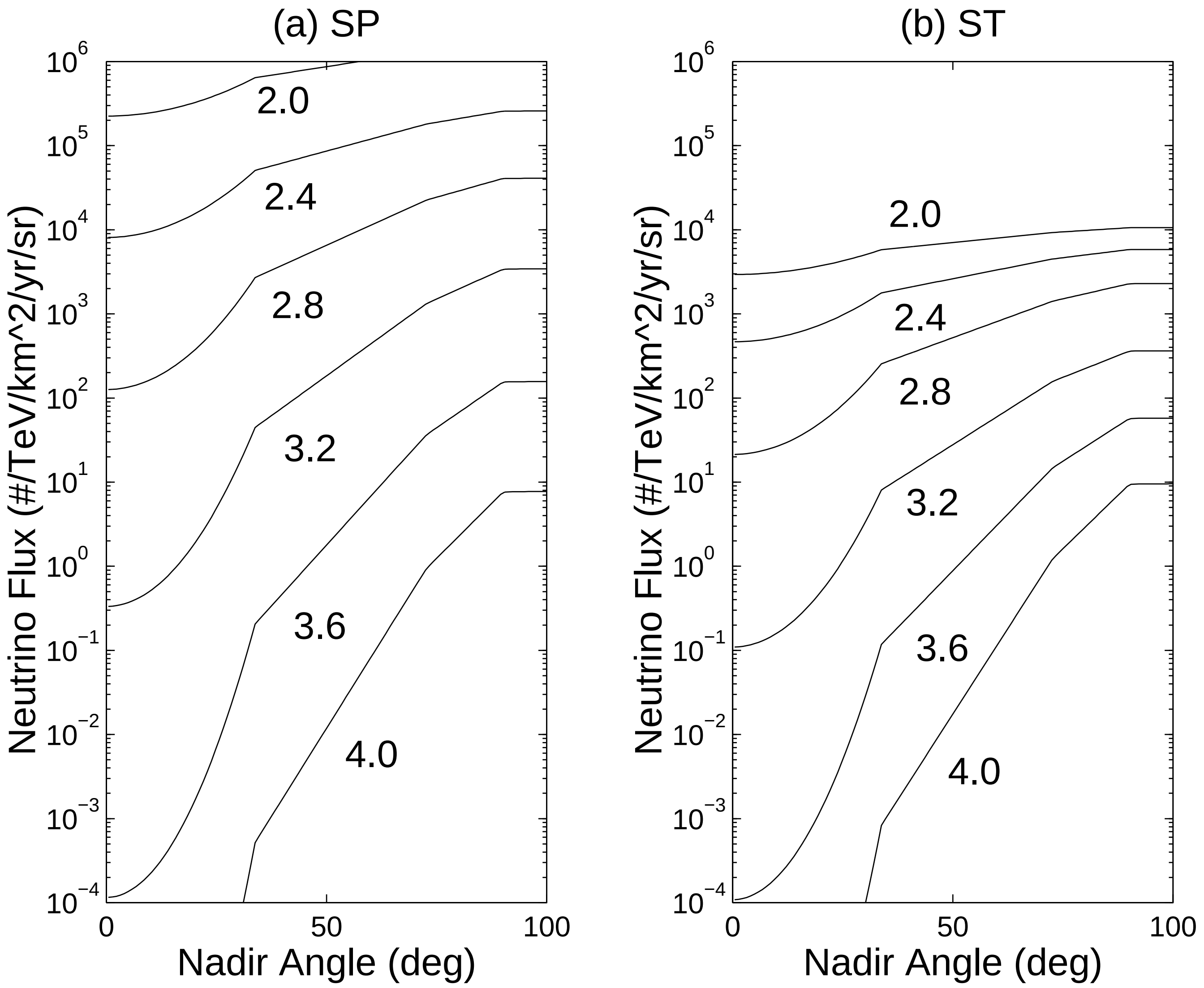

Figure 5: Diffuse AGN neutrino flux in the detection region showing calculated angular variation due to Earth shadowing. Neutrino flux versus angle with respect to vertical downward is plotted for a series of neutrino energies from 100 TeV to 10 PeV. Curves are labeled with $\log_{10}$[energy(TeV)]. Parts (a) and (b) show the total flux of $\nu_e$, $\bar{\nu}_e$, $\nu_\mu$, and $\bar{\nu}_\mu$ using the SP and ST predictions respectively.

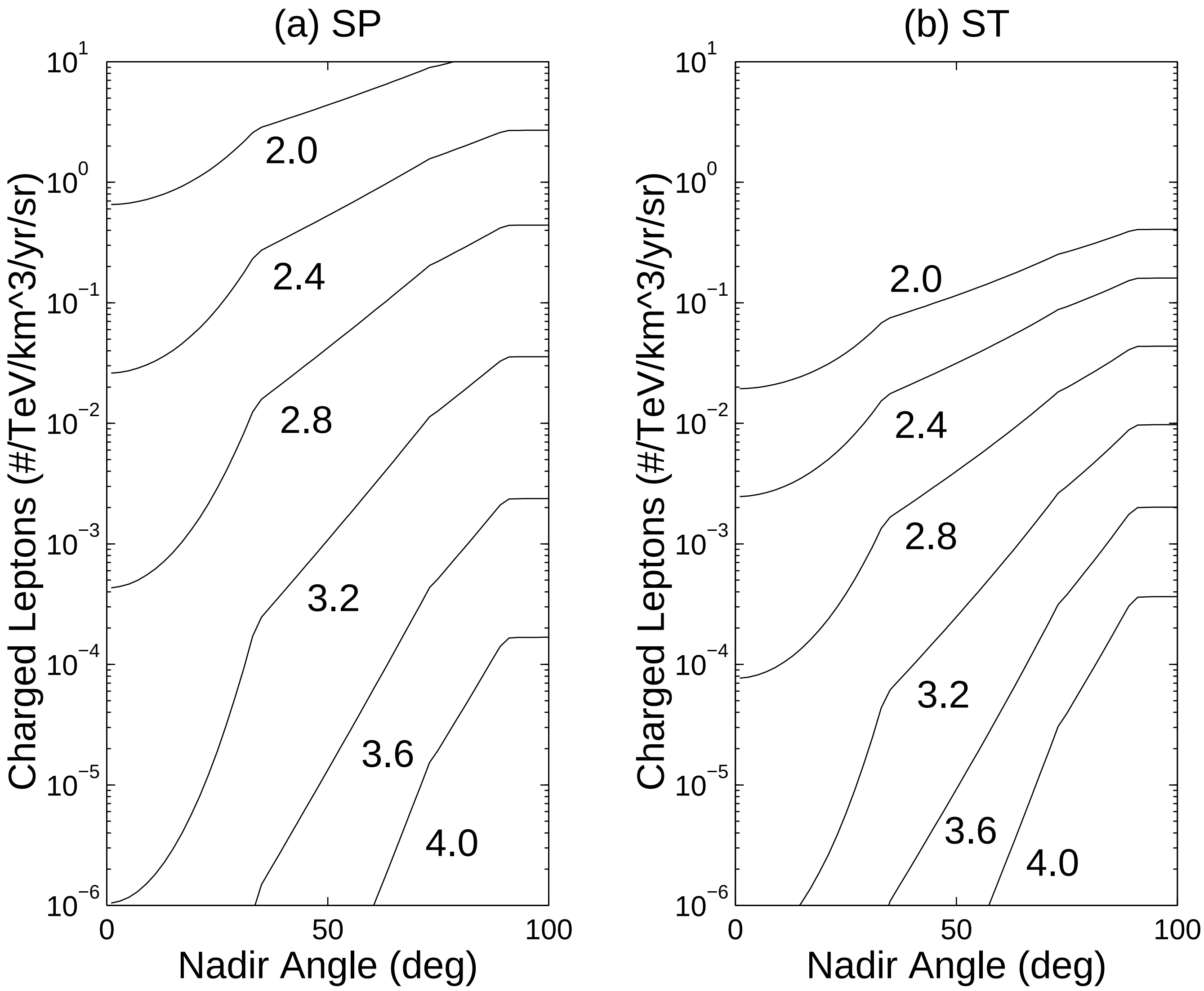

Figure 6: Charged lepton production ($\mu^{\pm}$,$e^{\pm}$) per year per TeV per steradian per cubic kilometer of polar ice due to charged current AGN neutrino interactions. The production rate versus angle with respect to vertical downward is plotted for a series of neutrino energies from 100 TeV to 10 PeV. Curves are labeled with $\log_{10}$[energy(TeV)]. Parts (a) and (b) show the rates induced by the SP and ST flux predictions respectively.

section.

# 5 Electromagnetic Shower Rates

We will present detailed calculations for radio detection of the process $\nu_e + p \to e + X$, followed by an electromagnetic shower caused by the energetic final state electron. We ignore radio energy produced by the hadronic shower, whose evolution involves strong interactions. It follows that our calculations of radio power and event rates are very conservative. Provorov and Zheleznykh [15], for example, treat hadron showers on the same footing as electron showers. The radio power from $\nu_e$ induced showers is nearly doubled by this effect, and $\nu_\mu$ events produce a full electron-like shower. Additionally, neutral-current neutrino interactions also produce observable showers. We will return to this point in the final section of this paper.

Contrary to optical detection, we find that electrons are more efficiently detected with radio than muons. This fact holds despite the long range of the muon. Every electron produces a shower carrying all of its energy, and whose compact evolution produces a sizable radio signal. Thus we have for electron showers the simple relation $\Gamma_e^{\rm shower}(E_s,\theta) = \Gamma_e(E_e,\theta)$. On the other hand, it is rare for a muon to bremsstrahlung a photon carrying a sizable fraction of its energy which would make for efficient radio detection. The small probability for bremsstrahlung closely compensates the effects of the muon's long range. Moreover, for every photon carrying a fraction $x$ of the muon's energy that will produce a shower carrying energy $E_s$, the muon has to have energy $E_\mu = E_s/x$. The distribution over $x$ tends to force $E_\mu$ up into a region of higher energy and lower flux, decreasing the rate.

To be more quantitative, we adopt a continuous loss expression for the muon shower rate $\Gamma_\mu^{\rm shower}(E_s,\theta)$ for showers moving at angle $\theta$ with respect to the nadir, given by

$$\Gamma_\mu^{\rm shower}(E_s,\theta) = \int_{E_s}^{\infty} dE_\mu \rho_{\rm ice} \frac{dP_B}{dE_\mu}(E_\mu, E_s) \int_0^{\mathcal{X}(\theta)} dr \Gamma_\mu(E'_\mu,\theta)\,, \tag{7}$$

with

$$E'_\mu = (E_\mu + \epsilon) e^{r/\xi} - \epsilon\,. \tag{8}$$

where $\epsilon$ and $\xi$ are the usual energy loss parameters, and $\frac{dP_B}{dE_\mu}$ is the bremsstrahlung probability [29]. The range integration of Eq.(7) finds the muon flux at en-ergy $E_\mu$ arising from muons produced at range $r$ and having an initial energy $E'_\mu$ given by the range-energy relation Eq.(8). The upper integration limit, $\mathcal{X}(\theta)$, is the chord length from the detector site to the Earth's surface for a fixed direction $\theta$.

Figure (7) shows our result for the number of EM showers occuring per year, per energy bin, per unit solid angle, per unit volume due to muon bremsstrahlung events. The curves have an interesting shape which peaks in the horizontal direction. This is to be expected since upward going muons are few in number due to neutrino absorption by the Earth, and smaller target volumes above the detector produce fewer muons compared with the more favorable horizontal direction. Recalling from Section 2 that muon neutrinos outnumber electron neutrinos by 3:2 in the SP case and 2:1 in the ST case, we nevertheless find that the peak horizontal shower rate is a bit more than a factor of 10 lower than the rate induced by the electron neutrino interactions. Given that the contribution of muons is small, the continuous energy loss expression, which neglects stochastic fluctuations of photon emission, is quite adequate for our study.

We now want to determine the number of these showers giving rise to a detectable signal in a radio receiver located some distance below the ice surface. This task involves several steps that will be dealt with in subsequent sections. Our calculation takes into account the details of EM shower structure in ice, a characterization of the resulting radio pulse, pulse attenuation as it propagates from interaction site to antenna, the coupling of signal to antenna voltage including polarization effects, the receiver characteristics transforming the induced antenna voltage into a detectable signal, and finally an estimate of the noise levels with which the desired signal from neutrino interactions must compete.

In the sections which follow, we will fold together what is known about each of these items and arrive at a conservative estimate for detectible neutrino events. This will then be expressed in terms of an energy/direction-dependent $(E, \theta)$ sensitive volume for an individual sensor (antenna plus receiver). This effective volume can be directly multiplied by $\Gamma_e^{\rm shower}$ and $\Gamma_\mu^{\rm shower}$ to arrive at an event rate predictions for electron and muon neutrinos respectively. In this way threshold effects are conveniently incorporated into the overall rate estimate.

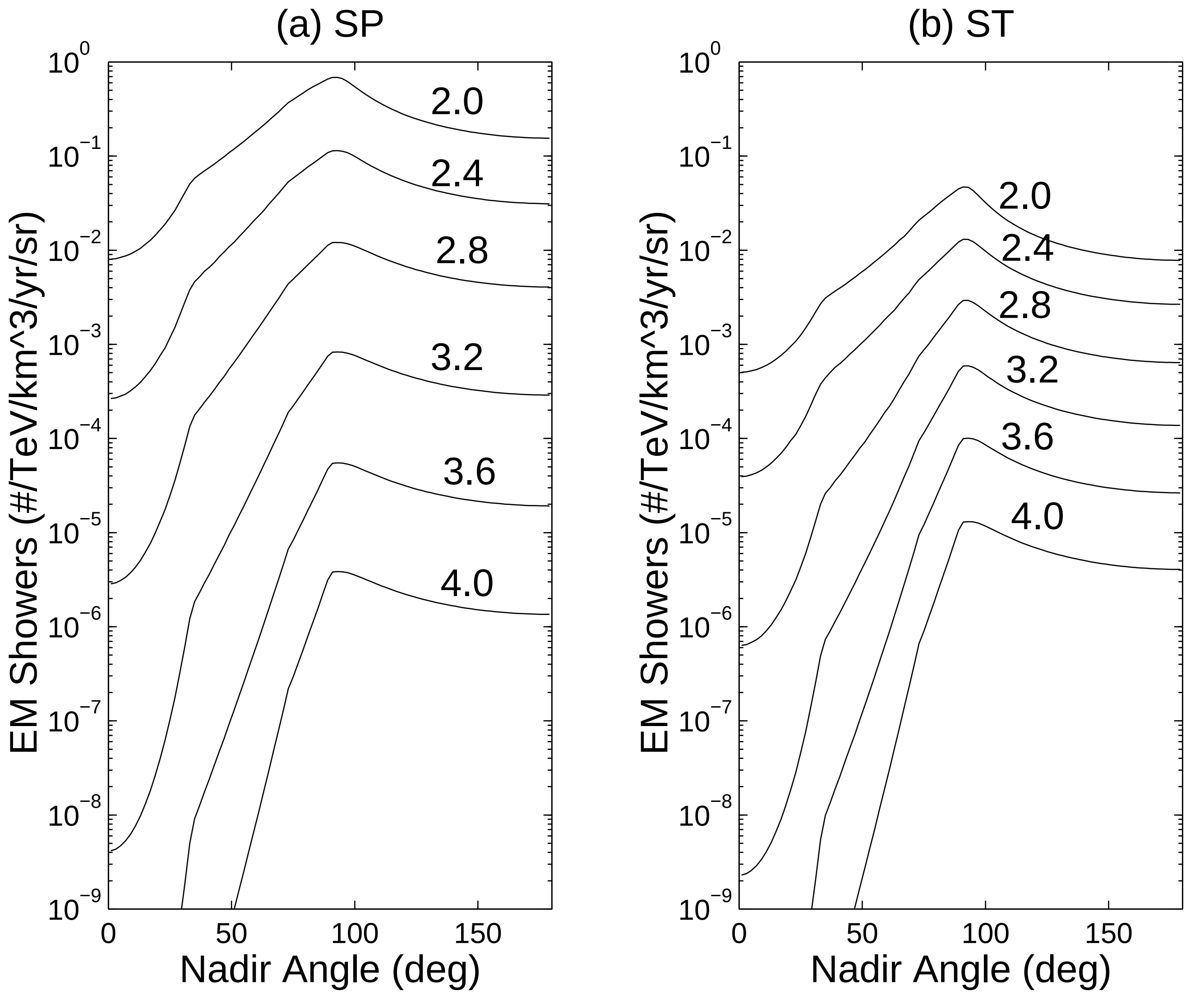

Figure 7: Number of EM showers induced by muon bremsstrahlung events per year per TeV per steradian per cubic kilometer of polar ice. The production rate versus angle with respect to vertical downward is plotted for a series of neutrino energies from 100 TeV to 10 PeV. Curves are labeled with $\log_{10}$[energy(TeV)]. Parts (a) and (b) show the rates induced by the SP and ST flux predictions respectively.

# 6 The Cherenkov Pulse and Its Propagation in Antarctic Ice

Scaling laws governing the overall behavior of radio detection have been understood for some time now [9] based on analytic models of shower structure together with the characteristics of Cherenkov emission. Recently, a significant advance in our understanding of coherent radio pulses from showers in ice was due to the detailed numerical study of Zas, Halzen and Stanev [14]. Their Monte Carlo code incorporates all the relevant particle interactions, tracks the history of each shower particle and then coherently sums the low frequency Cherenkov radiation from all the resulting charged particle tracks. The simulations give a very detailed view of the shower as it propagates through the ice with its accompanying radio pulse.

The relevant result for our purposes is the variation of the signal frequency spectrum with primary shower energy and also the angle between the longitudinal shower axis and the observation direction. Near the Cherenkov angle ($\theta_c = 56^o$ in ice), the Fourier transform of the pulse electric field, $\vec{E}(\nu, R, \theta_c)$, can be parameterized compactly as follows [14],

$$R|\vec{E}(\nu, R, \theta_c)| = \frac{0.55\times 10^{-7}(\nu/\nu_o)}{1+0.4(\nu/\nu_o)^2}\frac{E_s}{1\mathrm{TeV}}\exp[-\frac{1}{2}(\frac{\theta-\theta_c}{\theta_o}\frac{\nu}{\nu_o})^2](\frac{\mathrm{V}}{\mathrm{MHz}}), \quad (9)$$

where R is the distance between observer and cascade, $\nu_o$ = 500 MHz, $\theta_o \sim 2.4^o$, and $E_s$ is the total shower energy. The deliverable power in this signal is proportional to the square of the electric field and scales as the primary energy squared with a peak near 1 GHz. This $E_s^2$ growth of the available signal power is to be contrasted with an $E_s$ dependence for incoherent emission at higher frequencies.

The next step is to determine how interaction with the ice modifies the pulse defined in Eq.(9) as it propagates from event site to antenna. Below about 5 GHz, dispersive effects on the pulse can be neglected, leaving absorption of pulse energy as the relevant effect to consider. The absorption is frequency and temperature dependent. We use the ice temperature versus depth data taken at the Antarctic station *Vostok* by Salamantin *et al.* [30] along with measurements of attenuation length variations with frequency and temperature due to Bogorodsky and Gavrilo [17] to predict the pulse spectrum after propagation along a given path from event to receiver.

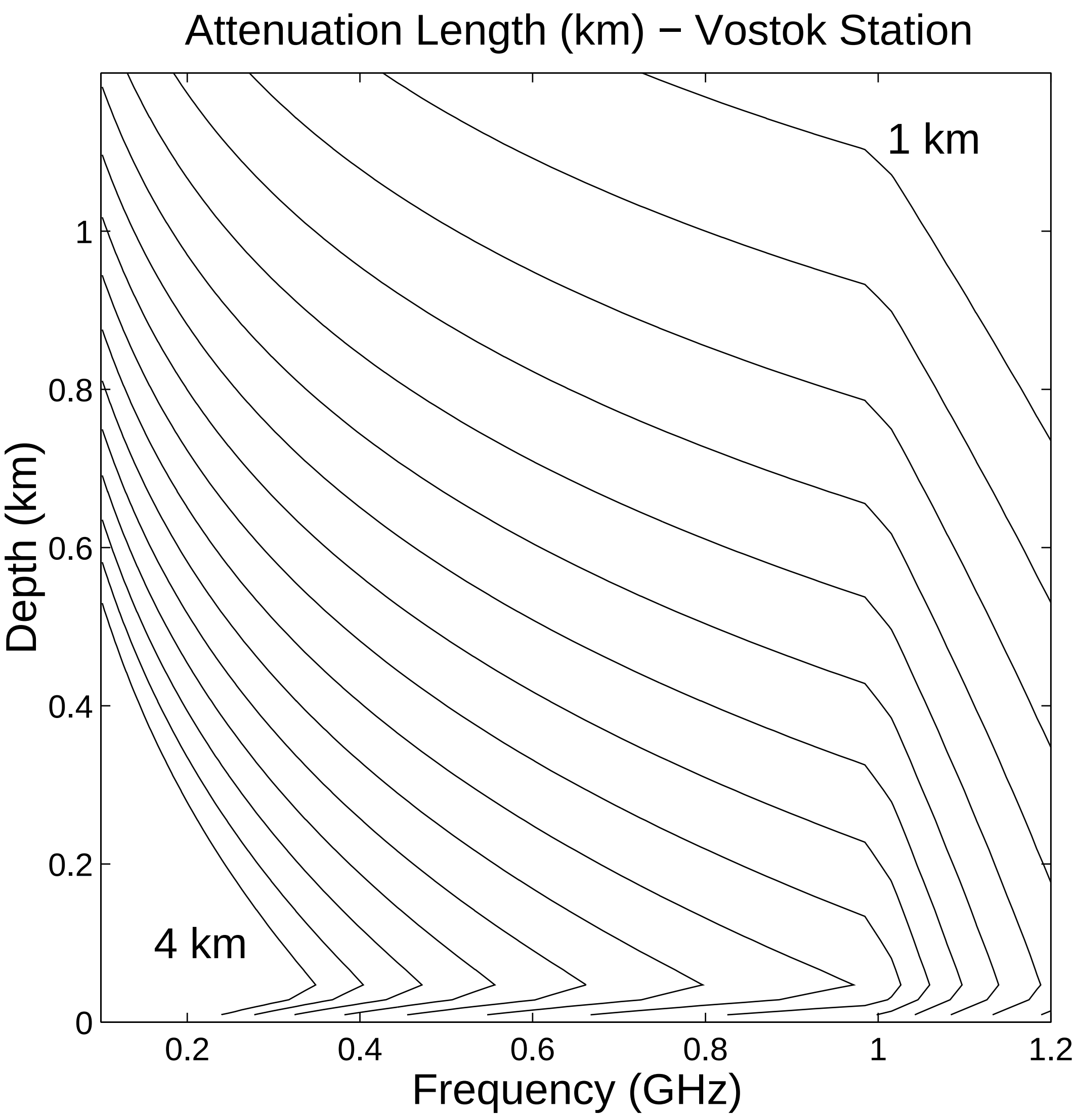

Figure 8: Expected absorption length in kilometers versus depth and frequency based on Antarctic temperature measurements at the *Vostok* station [ref. 30]. Contours are in increments of 0.2 km from 1 to 4 km.

In Figure (8) we show the resulting attenuation length as a function of frequency and depth below the ice surface. At 400 meters depth, the absorption length ranges from 4.4 km at 100 MHz to 2.0 km at 1 GHz. Just above 1 GHz new loss mechanisms become active and the attenuation increases sharply. Because it has the lowest temperature, the ice nearest the surface is the most radio-transparent. By itself, this factor would suggest shallow deployment of a radio receiver for optimum event detection. However, recall from the results of Section 3 that most of the neutrino flux is coming from the upper hemisphere due to the phenomenon of shadowing by the Earth. This means that there are advantages to having a large detection volume above the receiver: that is, deep deployment. The depth which yields the optimum overall rate of detectable events is a quantity that is vital for detector design, and we will evaluate it in subsequent sections.

As an example of signal attenuation, Fig.(9) shows attenuated spectra for pulses propagating a distance, $R = 2$ km, to a receiver located at a depth of 700 m from a series of five angles ranging from $30^o$ to $110^o$ measured with respect to vertical downward. The variation of the spectrum magnitude and corresponding pulse height with angle underscores the virtues of lower temperature layers and the advantages of operating the detector in the more radio-transparent ice.

# 7 Coupling of Signal to Detector

The attenuated pulse from the previous section arrives at the antenna having a particular polarization determined by the geometry of the event. The electric field, $\vec{E}$, is transverse to the propagation direction and lies in the plane containing the pulse's Poynting vector and the longitudinal shower axis. This field induces an open-circuit voltage, $V_a$, in the receiving antenna that can be expressed in terms of an effective height vector, $\vec{h} = h(f)\hat{\psi}$, which, together with antenna gain, expresses the efficiency of the antenna as a transducer between the propagation medium and a transmission line connected to the antenna terminals. The angular variable, $\psi$, refers here to the angle between the incoming signal's Poynting vector and the symmetry axis of the antenna. The effective height function depends on the material properties and geometry of the antenna itself as well as the dielectric properties of its environment, in this case the ice in which it is imbedded. The relation is

$$V_a = \sqrt{\epsilon}\vec{E}(f,\psi) \cdot \vec{h}(f)G(\psi)\,, \tag{10}$$

where $G(\psi)$ is the gain of an omnidirectional antenna, and $\sqrt{\epsilon} = 1.78$ is the refractive index of ice.

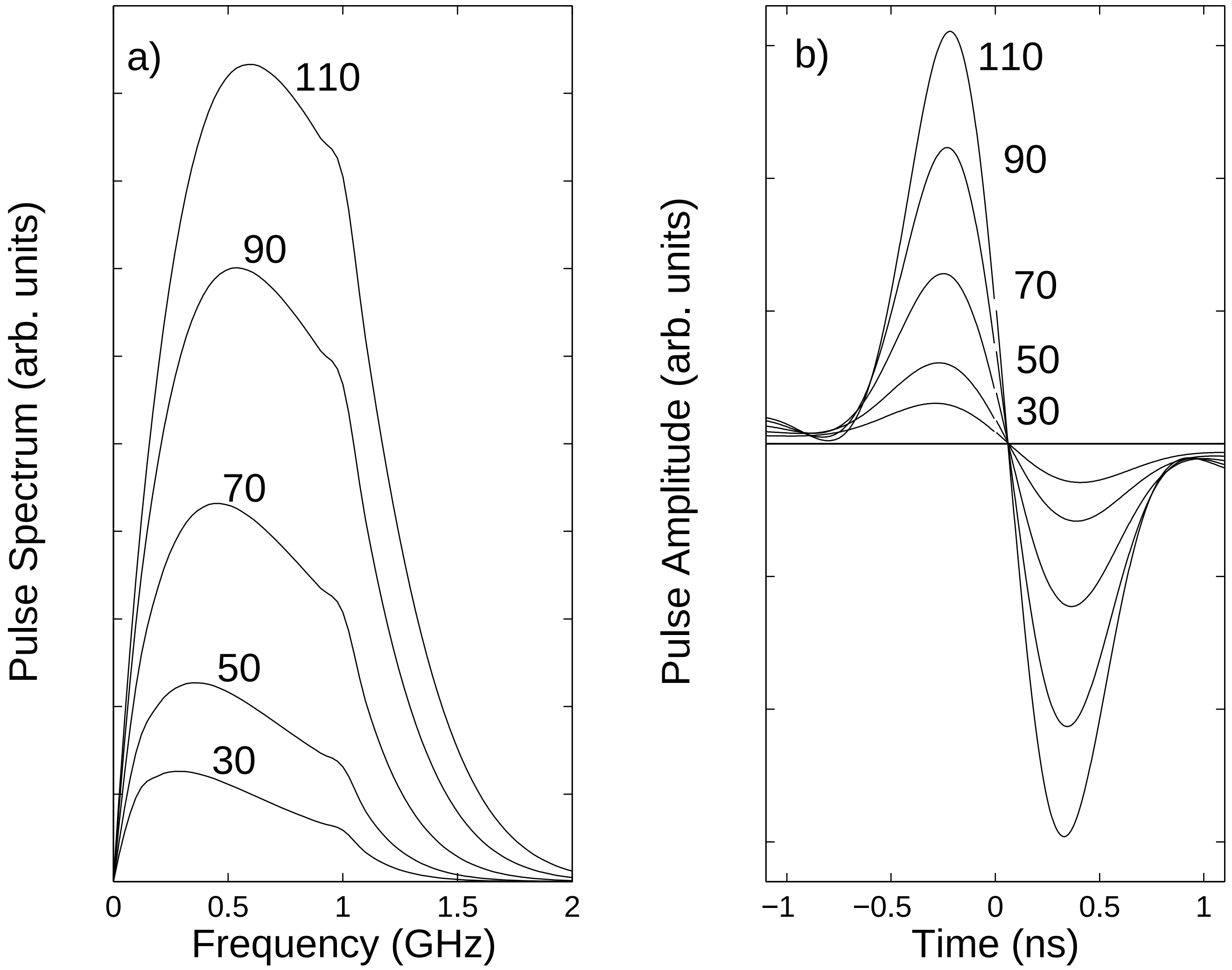

Figure 9: Cherenkov pulse after propagating 2000 meters from the cascade site to a receiver at 700 meters depth. The pulse is shown in the frequency domain (modulus only) and time domain with the curve labels indicating the nadir angle in degrees for the event. Pulses passing through the colder ice layers suffer less attenuation.

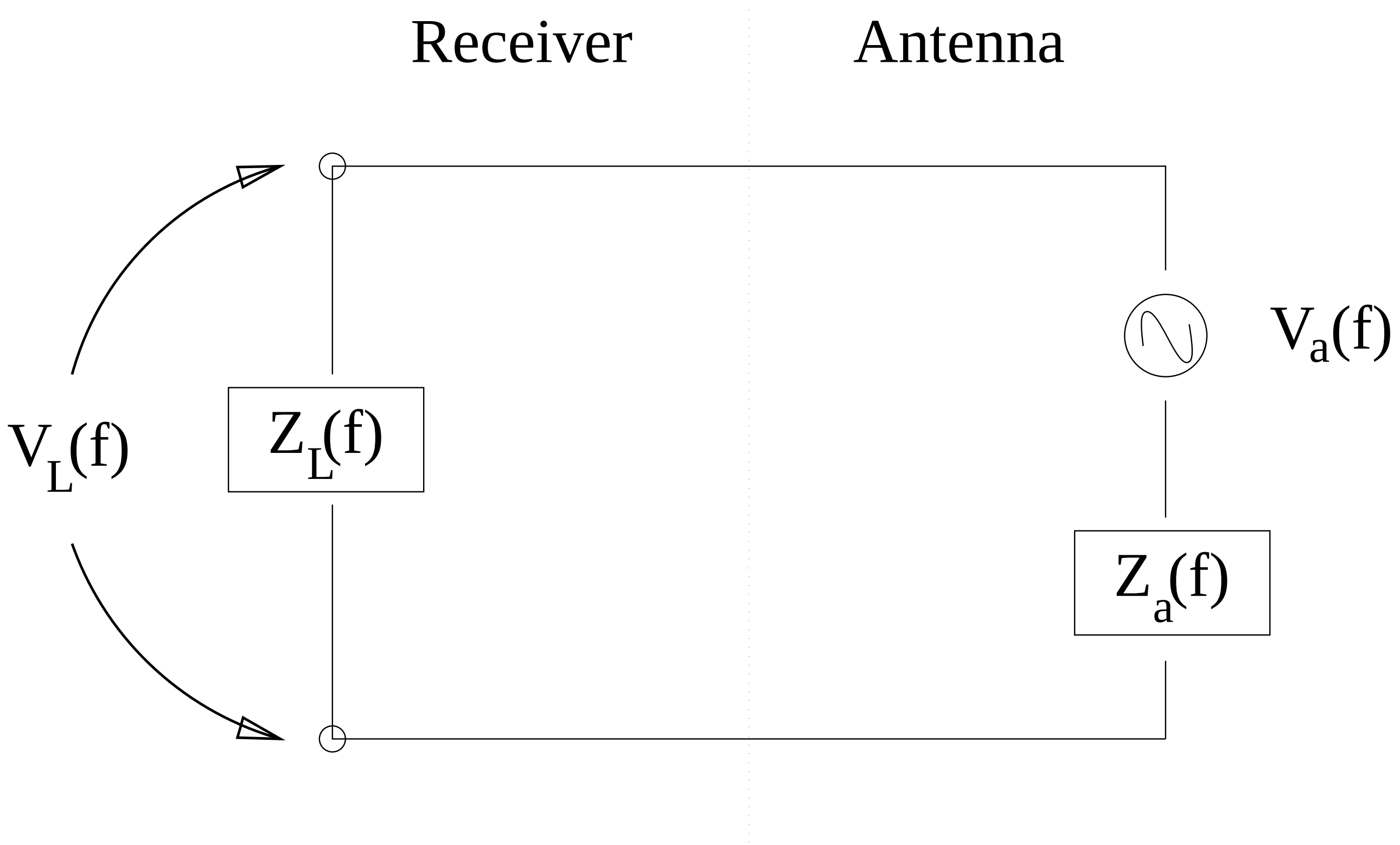

Figure 10: An equivalent circuit representing the antenna-receiver combination. These quantities will be used in the evaluation of noise and signal charateristics at the receiver output.

The voltage of Eq.(10) drives the equivalent receiver plus antenna circuit depicted in Fig.(10), where $Z_a(f)$ and $Z_L(f)$ represent the frequency dependent complex impedances of the antenna and the receiver load respectively. Under matched impedance conditions, the receiver is such that $Z_L(f) = Z_a^*(f)$ ensuring maximum power transfer from antenna to receiver. In this case, the signal voltage we measure across the receiver load is simply,

$$V_L = \frac{V_a Z_a^*}{2\mathrm{Re}(Z_a)} \,. \tag{11}$$

For our numerical work we will use theoretical predictions for the effective height, gain and impedance of a particular wide-angle biconical antenna. Expressions for these antenna characteristics can be found in references [31]. Existing impedance measurements for conical antennas [32] match the theoretical values quite well. We model a biconical antenna having a total length of 22 cm and a conical half-angle of $30^o$. Results are shown in Fig.(11). One can see from the effective height that the sensitivity peaks in a band from roughly 100 to 400 MHz and falls off sharply at higher frequencies. At first this lack of efficiency near 1 GHz might seem like a serious liability since the initial pulse spectrum of Eq.(9) peaks there. Recall, however, that the signal spectra shown in Fig.(9) peak at lower and lower frequency as attenuation is increased. This means that signals near the detection threshold are likely to have their power concentrated at frequencies where our biconical antennas are most sensitive. Most of our detectable events will come from large distances (an $r^3$ volume effect) and near threshold. From the point of view of maximizing the rate of detectable events, the sensitivity of the biconical described above seems to make it a good choice.

The biconical antenna should perform somewhat better than, for example, a simple dipole of the same length because of its larger bandwith, helping to discriminate the signal pulse from noise. It is also an attractive option because it has a maximum horizontal diameter of about 10 inches, a size already shown to be practical for placing down deep bore holes by the AMANDA collaboration.

The impedance and effective height functions of Fig.(11) are used in Eqs. (10) and (11) along with the attenuated pulse spectrum from Section 6 to determine the signal at the receiver output. As an example, Fig.(12) shows the spectrum at the receiver output, $V_L(f)$, for the five signals described in Fig.(9). This is the signal we wish to detect.

# 8 Signal to Noise Ratio

The total voltage observed at the receiver's output can be expressed as the sum of the signal given by Eq.(11) and a random thermal noise component.

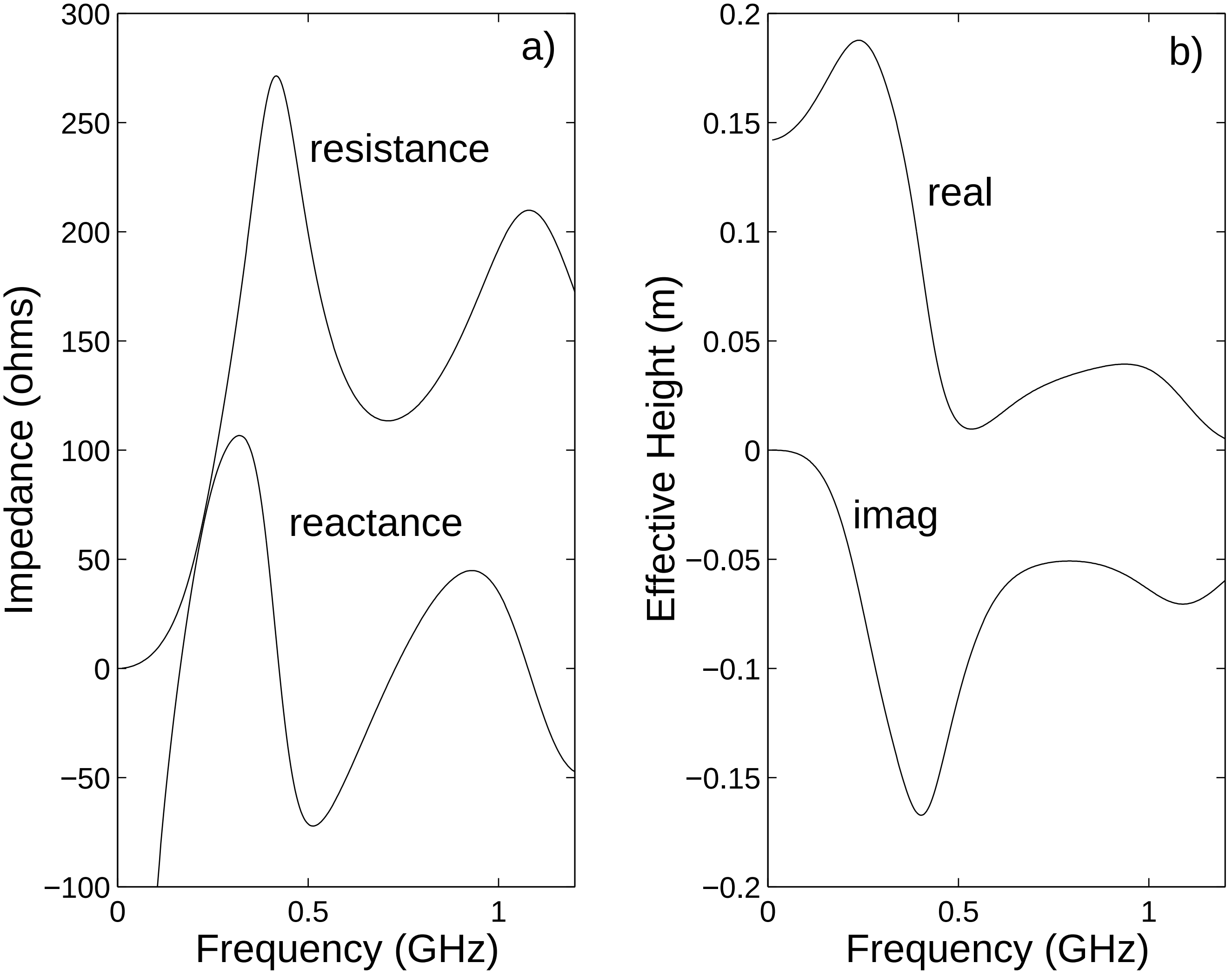

Figure 11: Antenna driving point impedance and effective height versus frequency. These are theoretical predictions for a $30^o$ (cone half-angle) biconical having a 22 cm total length.

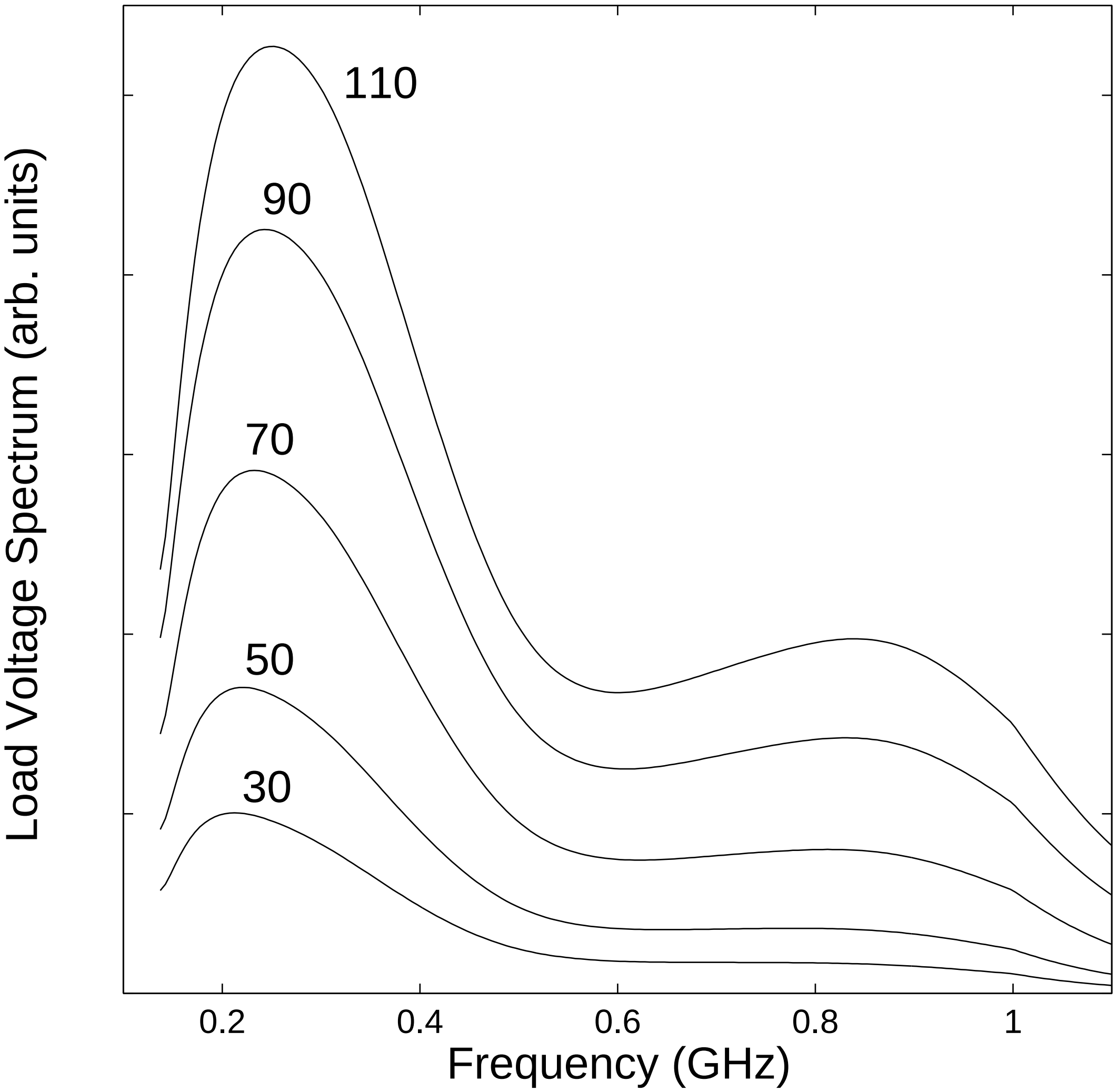

Figure 12: Magnitude of load voltage spectrum given by Eq.(11) corresponding to the five example signals of Fig.(5)

This includes Johnson noise generated in the circuitry of the receiver itself, as well as other random electromagetic fields present in the environment that are absorbed by the antenna. This type of noise is characterized by a flat spectrum whose magnitude can be expressed in terms of an equivalent noise temperature, $T_n$. The noise power in a band, $\Delta f$, is simply $kT_n\Delta f$. We assume that detection of the neutrino signal will be limited by thermal noise and that backgrounds such as radio emission from the galactic plane or man-made sources are not important. This assumption seems reasonable based on measurements of ambient radio noise in the Antarctic by Boldyrev *et al.* [33]. These measurements indicate that the value $T_n = 300$K is a reasonable estimate over the 100 MHz to 1 GHz band that we will be concerned with in this work.

Detection of a known signal in a white Gaussian noise background is a standard problem of signal processing and is optimized in the sense of maximizing the signal to noise ratio,

$$\mathrm{SNR} = \frac{\text{peak output signal power}}{\text{output noise power}}\,, \tag{12}$$

by use of a matched correlation receiver. The matched filter employs a receiver transfer function that is 'matched' to the complex conjugate of the expected receiver input voltage. In principle we know the form of the expected signal and can make use of the increased discrimination against thermal noise that this procedure permits. Under matched conditions, the signal to noise ratio can be expressed,

$$\mathrm{SNR} = \frac{2}{kT_n}\frac{(\int_{-\infty}^{+\infty}|V_L(f)|^2 df)^2}{\int_{-\infty}^{+\infty}|V_L(f)|^2 R_L(f) df}\,, \tag{13}$$

where $R_L$ is the resistive component of the load impedance, $Z_L$. A discussion of thermal noise, the matched filter and the associated expression for SNR can be found in reference [34]. We will consider detectable any event having SNR $\geq 1$. This is a very conservative method because use of correlations between different detectors can improve upon this considerably.

# 9 Effective Volume for the Matched Bi-Conical Antenna

Consider a fixed shower energy, $E_s$, incident neutrino direction, $\theta$, and a variable position for the electromagnetic shower relative to the antenna. The collection of potential shower positions that satisfys SNR $\geq 1$ using Eq.(13) defines an effective volume for events with fixed $(E_s, \theta)$. This volume can then be multiplied by $\Gamma_e^{\rm shower}$ or $\Gamma_\mu^{\rm shower}$ to directly give the rate for events initiated by neutrinos of electron and muon type respectively.

As an example, Fig.(13) depicts a vertical cross section of ice containing a receiver 900 meters deep. The contours indicate SNR=1 for shower energies of $E_s$ =1 PeV and 100 PeV with the incoming neutrino direction, $\theta = 90^o$, fixed (horizontal flux coming from the right side of the figure). At each energy, the two lobes (the upper lobe is cut off by the surface) are sections of the complete effective volume that can be imagined by rotating the lobes about an axis containing the receiver and pointing in the direction of the incident neutrino flux. This volume is simply an 'image' of the signal Cherenkov cone (with some distortions introduced by the variable properties of the ice with depth), truncated by its intersection with the ice surface at energies above about 1 PeV. This "fat" cone-shaped region is the effective volume of the receiver for that particular energy and flux direction. Each possible combination of $E_s$ and $\theta$ has its own cone-shaped volume. Note that at energies above 1 PeV the lobes extend more than a kilometer from the receiver. One could make similar sets of contours for other values of $\theta$ and receiver depth in order to see what sections of the ice are probed by a particular placement of the antenna. This procedure would be useful in designing a radio receiver array.

Figure (14) shows results for the effective volume of a 600 meter deep antenna versus nadir angle (the angle with respect to local downward vertical) and total cascade energy. Curves are drawn for a series of energies from 0.25 to 32 PeV. Considering the energy dependence of Eq.(9) along with the conical geometry of the sensitive region, one predicts a volume scaling like $E_s^3$ in the absence of attenuation. In practice, both the signal loss and the cutoff presented by the ice surface cause the volume to grow more slowly with energy; the volume scales like $E_s^3$ up to 200 TeV, $E_s^2$ near 1 PeV, and by 10 PeV the volume is increasing linearly with energy.

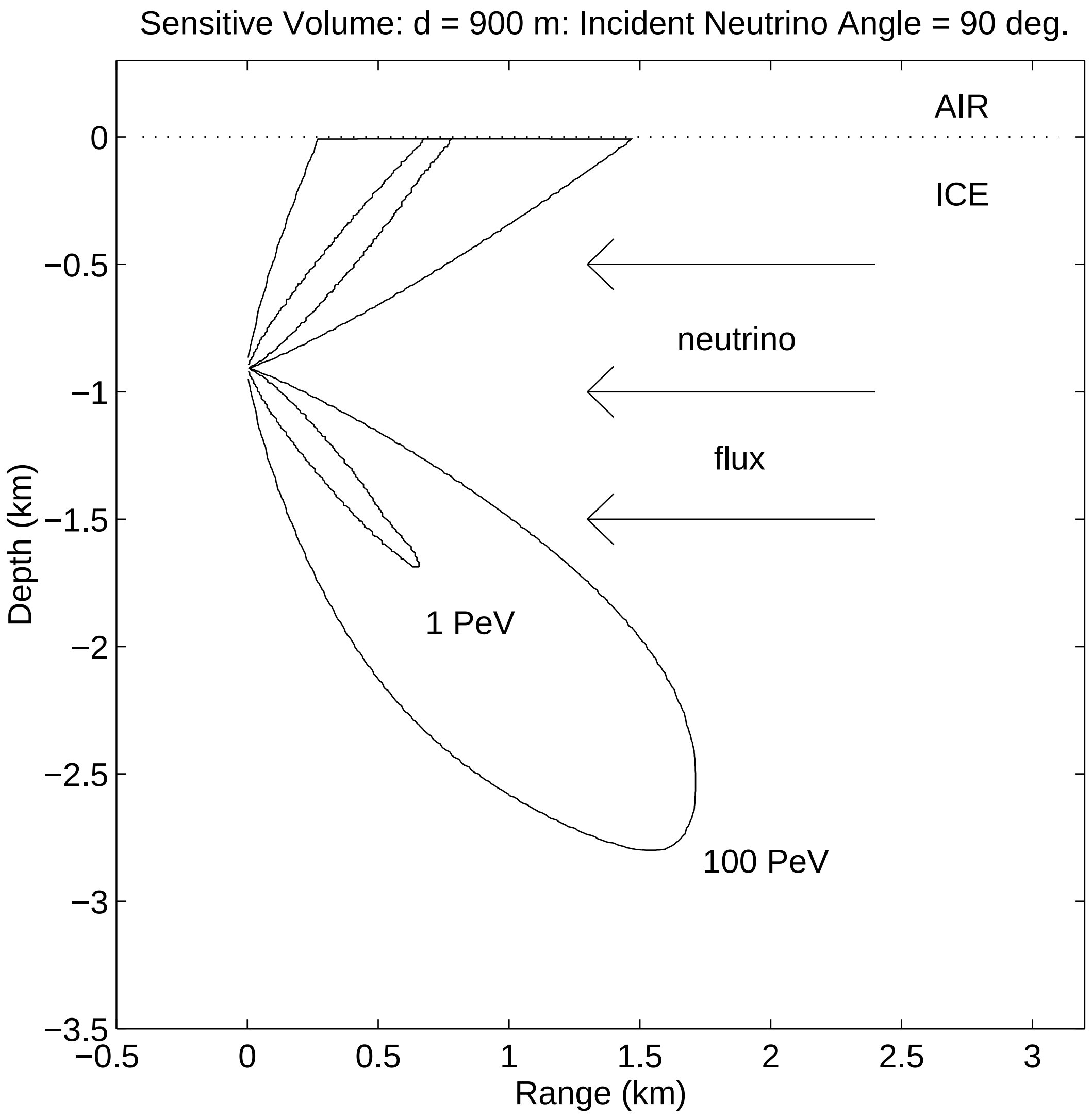

Figure 13: Visualization of the effective volume of an antenna located 900 meters below the ice surface. The two sets of contours outline the regions of SNR $\geq 1$ for shower energies of 1 and 100 PeV. At 1 PeV, the sensitive region already extends over a kilometer away from the antenna location. This volume covers different regions of ice for each possible incident neutrino direction. Here we show the situation for a horizontal flux incident from the right hand side of the figure.

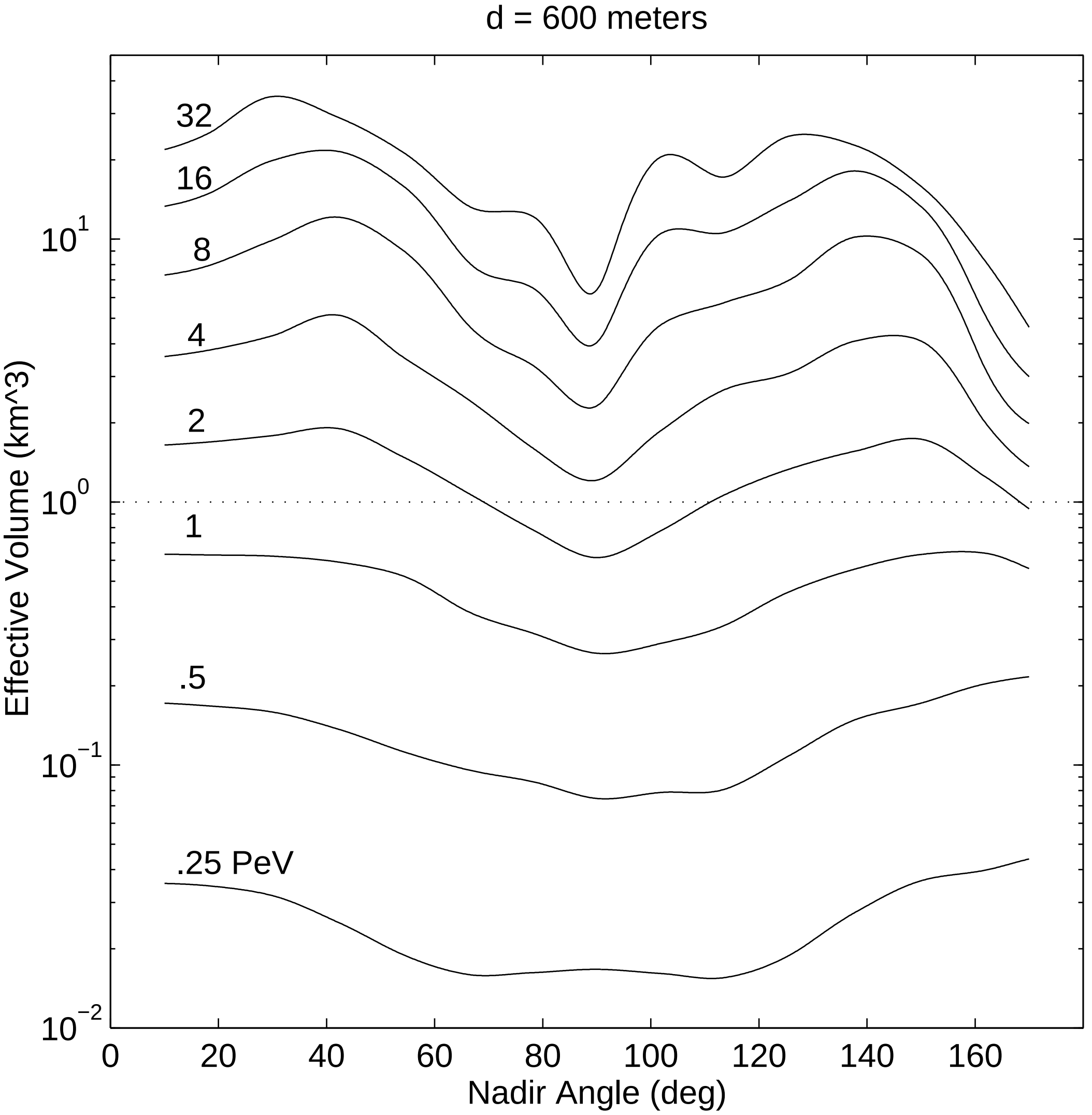

Figure 14: Effective volume for an antenna located 600 meters below the ice surface as a function of incoming neutrino direction and shower energy. Each curve is labeled with the event energy in PeV. The $1\mathrm{km}^3$ level is indicated by the dotted line.

At first glance, the angular dependence seen in Fig.(14) is unexpected. Based on antenna gain, one might guess that the antenna is most sensitive to the horizontal neutrino flux (*i.e.* angles $\sim 90^o$). Our results show a relative *minimum* for horizontal flux. The reason for this is the fact that the incident neutrino direction and that of the signal pulse differ by the $56^o$ Cherenkov angle in ice, hence the maxima at roughly $90^o + \theta_c$ and $90^o - \theta_c$. A more subtle contribution to this same effect is the favorable relative polarization of the signal's electric field and the antenna height for angles significantly above and below horizontal.

The depth dependence can be seen in Fig.(15), where we plot the effective volume versus nadir angle for $E_s = 5$ PeV and four depths from 100 and 1300 meters. Recalling that the ice is becoming increasingly radio-opaque with depth, and the fact that shallower deployments will be more severely affected by the surface cutoff, one can readily understand the qualitative behavior seen in this figure. Because of Earth shadowing of the incoming flux, we want to optimize the sensitivity over the upper hemishpere. It is not clear from this figure which depth will maximize the number of observed events. In the next section we will look at event rates versus antenna depth.

The important result seen in Figs.(13) through (15) is that sensitive volumes of order $1\text{km}^3$ are possible with as few as 1 or 2 receivers at 1 to 2 PeV, and roughly 10 receivers at 500 TeV. These numbers indicate that unprecedented target masses for UHE detection are feasible using radio as the detection mechanism. This is precisely what is required if the currently predicted UHE fluxes from AGN are to be observed. In the next section, we will fold in the expected cascade rates based on model AGN fluxes and arrive at definite rate predictions.

# 10 Event Rates

We now have in place all the elements to arrive at event rates for one of our proposed antennas. The effective volumes given in Section 9 can be combined with the EM shower rates presented in Section 5 to give a prediction for the number of events induced by electron and muon neutrinos. These rates are functions of the event energy and the incident flux angle and can be integrated to give a total rate. This integrated rate for the SP and ST AGN flux predictions is shown in Fig.(16) as a function of antenna depth.

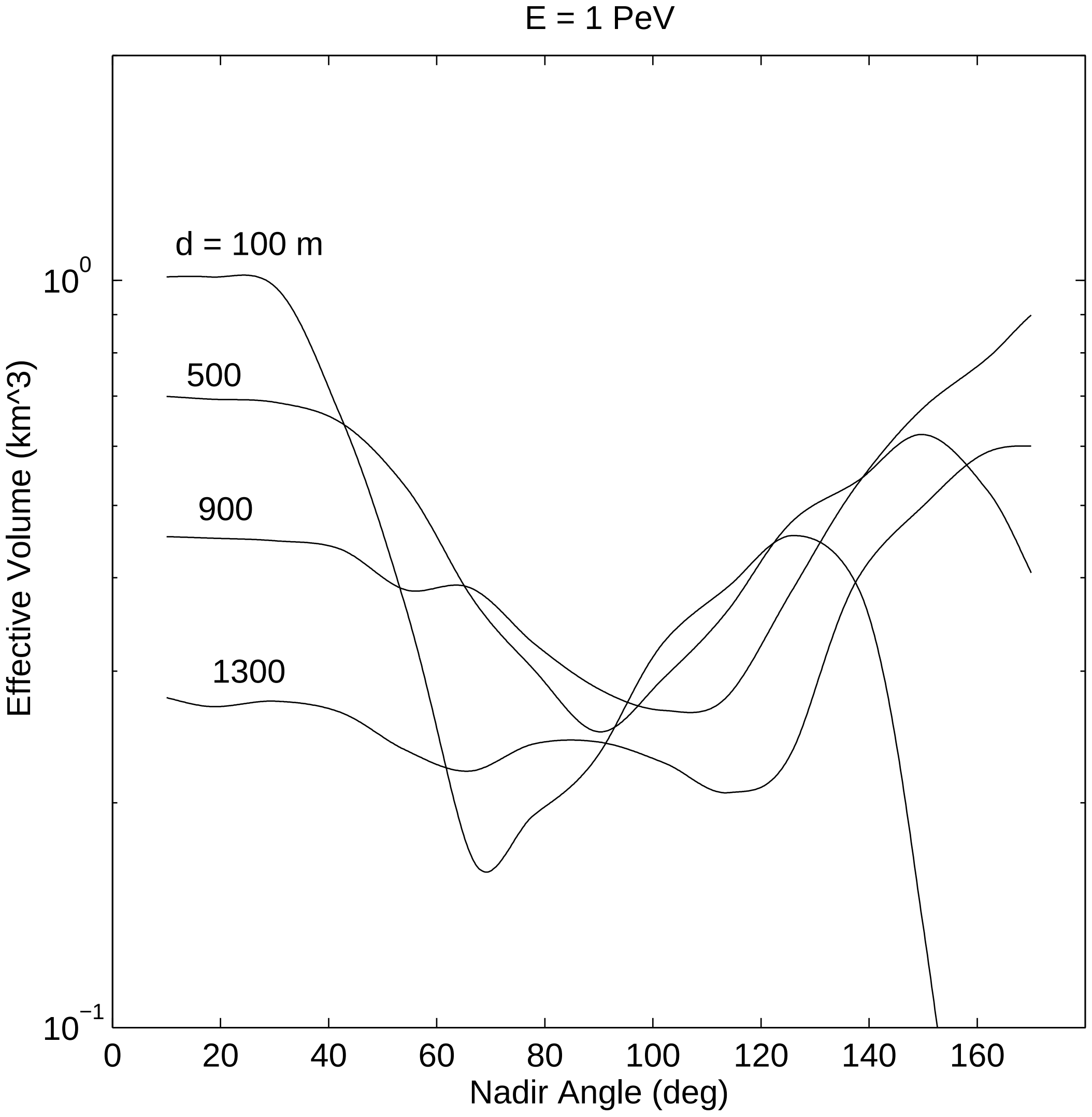

Figure 15: Effective volume for observing events having an energy of 1 PeV as a function of incoming neutrino direction and antenna depth. Each curve is labeled with the receiver depth in meters.

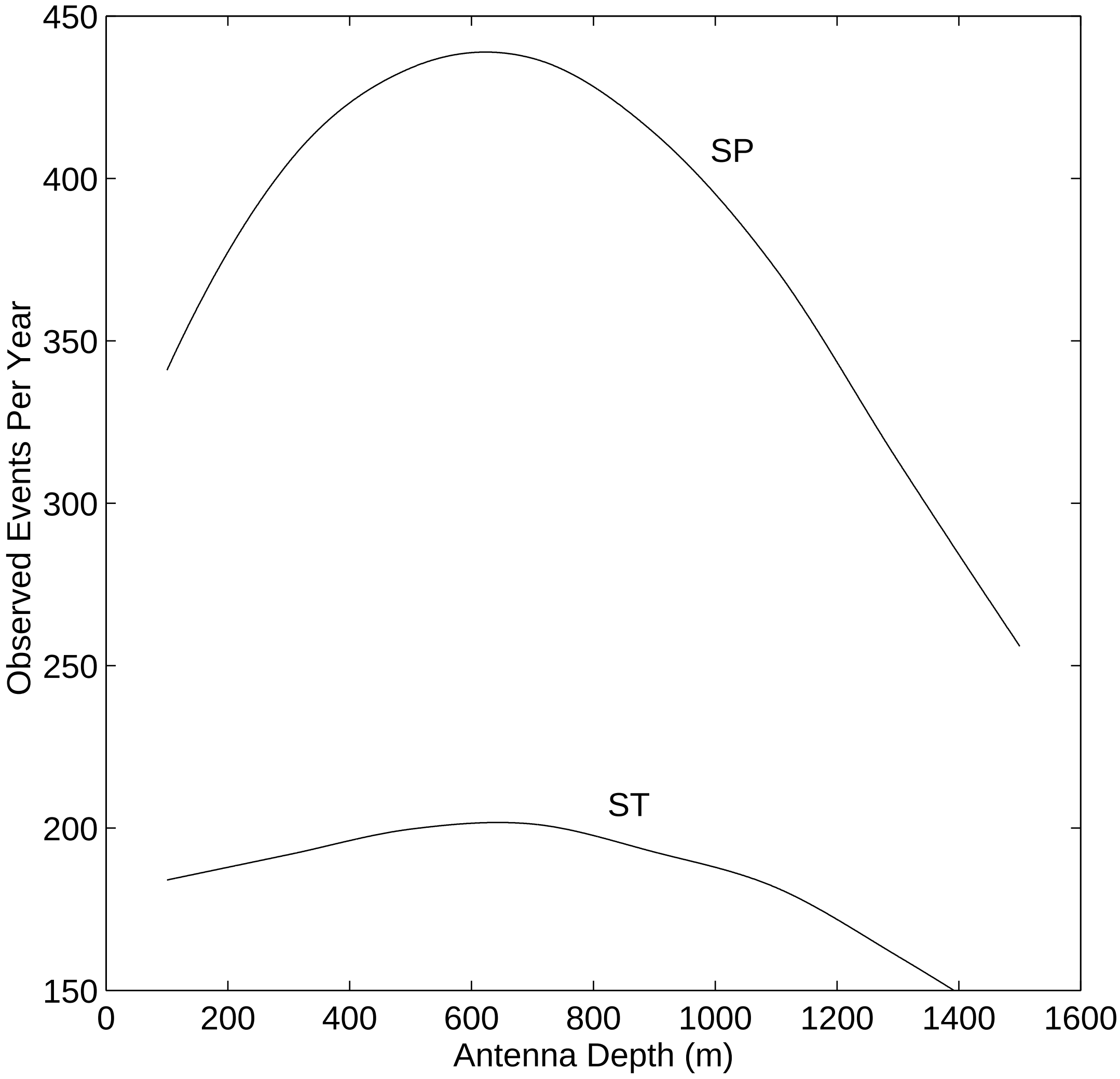

Figure 16: Total number of observed events by a single antenna per year versus antenna depth for the SP and ST AGN flux predictions. The optimum depth is about 600 meters.

We emphasize that the rates shown are for a *single* antenna. Any realistic deployment, even in the case of a small pilot experiment, would likely employ an array of such sensors. The overall rate is maximized for an antenna depth near 600 meters. The rate per antenna is 437 events per year for the SP flux and 201 events per year for the ST prediction. The result using the AT flux as input is a completely negligible 0.06 events per year. Clearly atmosperic neutrinos are not a background for radio detection of AGN neutrinos.

Figure (17) shows the logarithmic energy dependence of the event rate (see caption) for a fixed incident neutrino direction of 140 degrees. It is the product of a rising effective volume and the falling EM shower rate, also shown. Taking the curves corresponding to the SP prediction as an example, one sees that below 1 PeV, the detector volume is rising fast enough to 'beat' the falling shower rate resulting in an increasing rate for observed events. Beyond the 1 PeV mark, the volume rises more slowly, the shower rate becomes more steep, and as a result, the observed rate peaks near 1 PeV and then begins to decrease. In the ST case, the peak occurs near 10 PeV because of the flatter shower spectrum.

The energy-angle dependence of the AGN rates are shown in Figs.(18) and (19) for an antenna depth of 700 meters. The contours indicate the concentration of events coming from about 50 degrees above horizontal and centered near 1 PeV in the SP case and 10 PeV in the ST case. In the next section we discuss how the primary neutrino spectrum can be estimated from the location of this peak in the distribution.

# 11 Spectrum Estimation

A full scale Antarctic experiment would likely employ on the order of 100 radio receivers arrayed along several vertical strings. Precisely how the antennas are arranged spatially is a detailed experimental question that will likely depend strongly on which measurements one wants to optimize. Assume for the moment that data is collected from an experiment in which the angle and energy of each event can be determined with enough precision to begin forming an image of the distribution shown in Figs.(18) or (19). According to the current flux estimates, many thousands of events might be collected yielding a reliable picture of the distribution peak as well as its rate of falloff in the $(E_s, \theta)$ plane. In this case, it would be a simple matter to unfold the effective detection volume, $V(E_s, \theta)$, from the data directly giving estimates of the AGN neutrino spectrum (horizontal flux, no Earth shadowing) and its attenuation by interactions in the Earth (angles below horizontal). The neutrino spectrum is critical to understanding AGN as particle accelerators. Important information on neutrino cross section and nucleon structure could come from the observed attenuation rate in the upper layers of the Earths crust (where target densities are known with some precision).

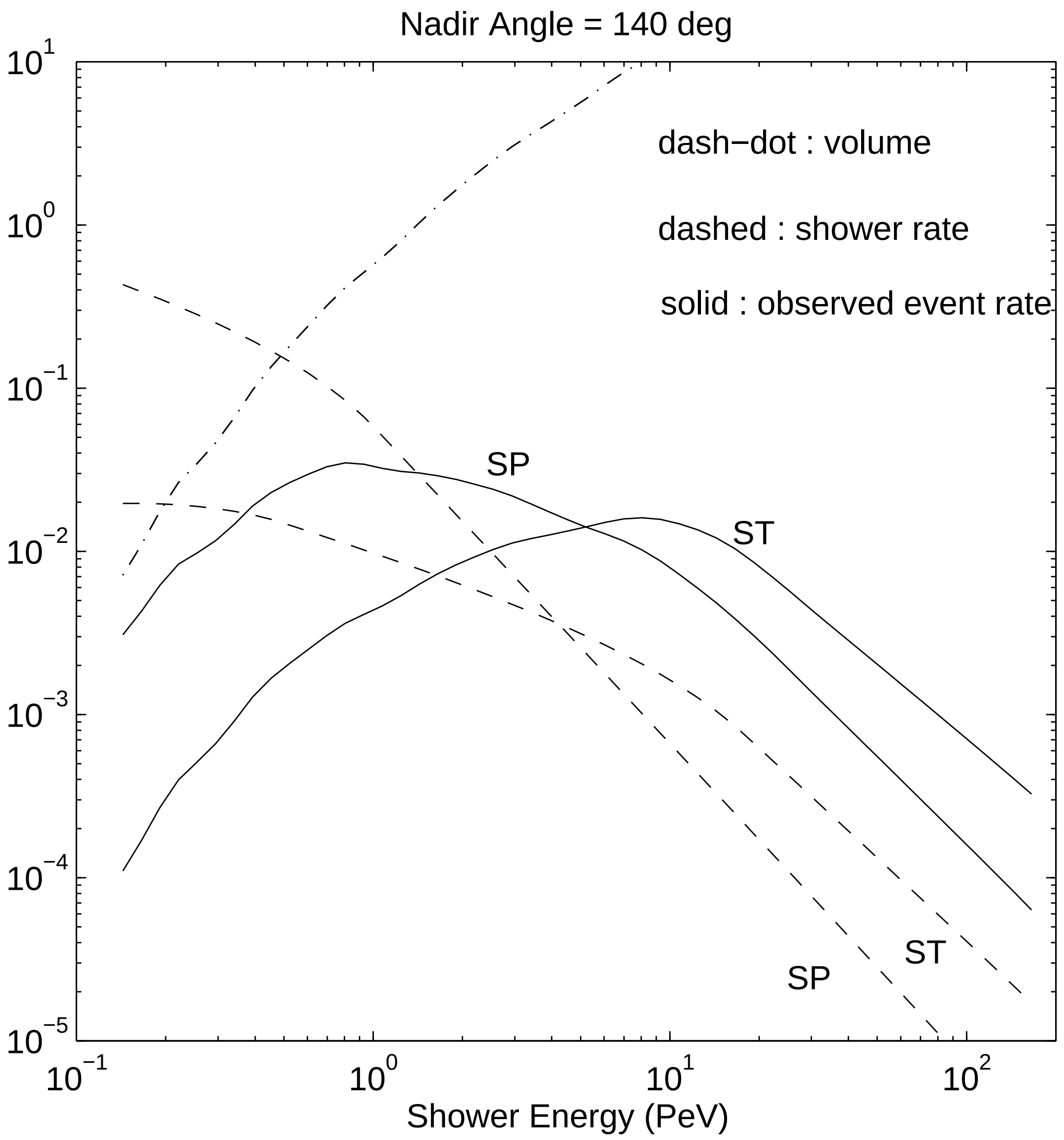

Figure 17: The event rate (solid line) per year is the product of the effective volume (dash-dotted line) in $km^3$ and the EM shower rate (dashed lines) per year, per $km^3$, per steradian, per Log(energy). We show results corresponding to an incident neutrino direction of 140 degrees with respect to the nadir for the SP and ST flux predictions. The difference in spectral indices of the SP and ST cases results in events peaking at 1 and 10 PeV repectively.

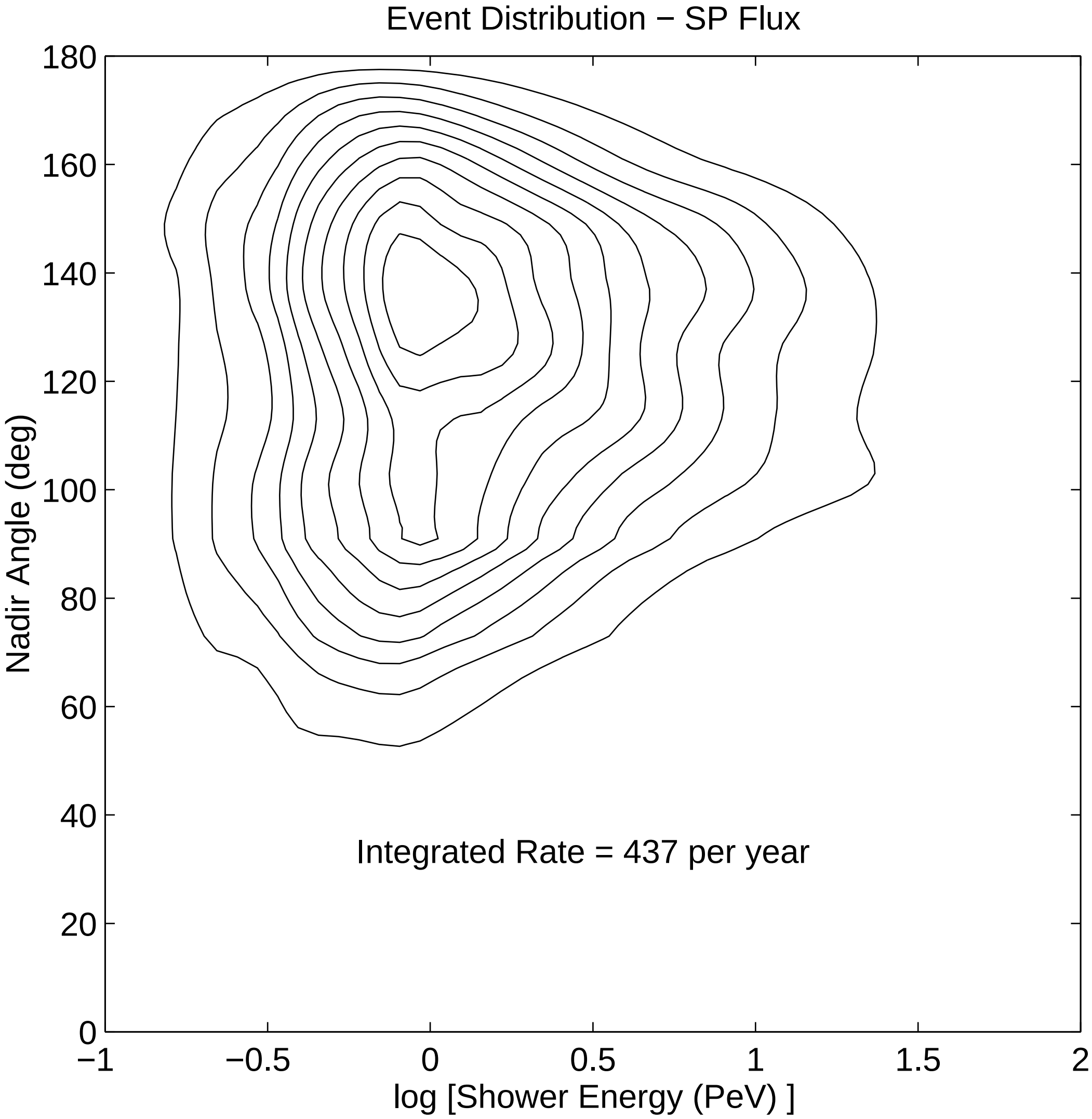

Figure 18: The distribution of AGN neutrino events as detected by a single antenna located 700 meters below the ice surface using the SP flux prediction. We plot 10 equally spaced contours of the quantity $dN/d\cos\theta/d\log E$, with the ratio between the innermost and outermost contours equal to 10. Events are concentrated about an energy of 1 PeV and 140 degree nadir angle. The integrated rate is 437 events per year.

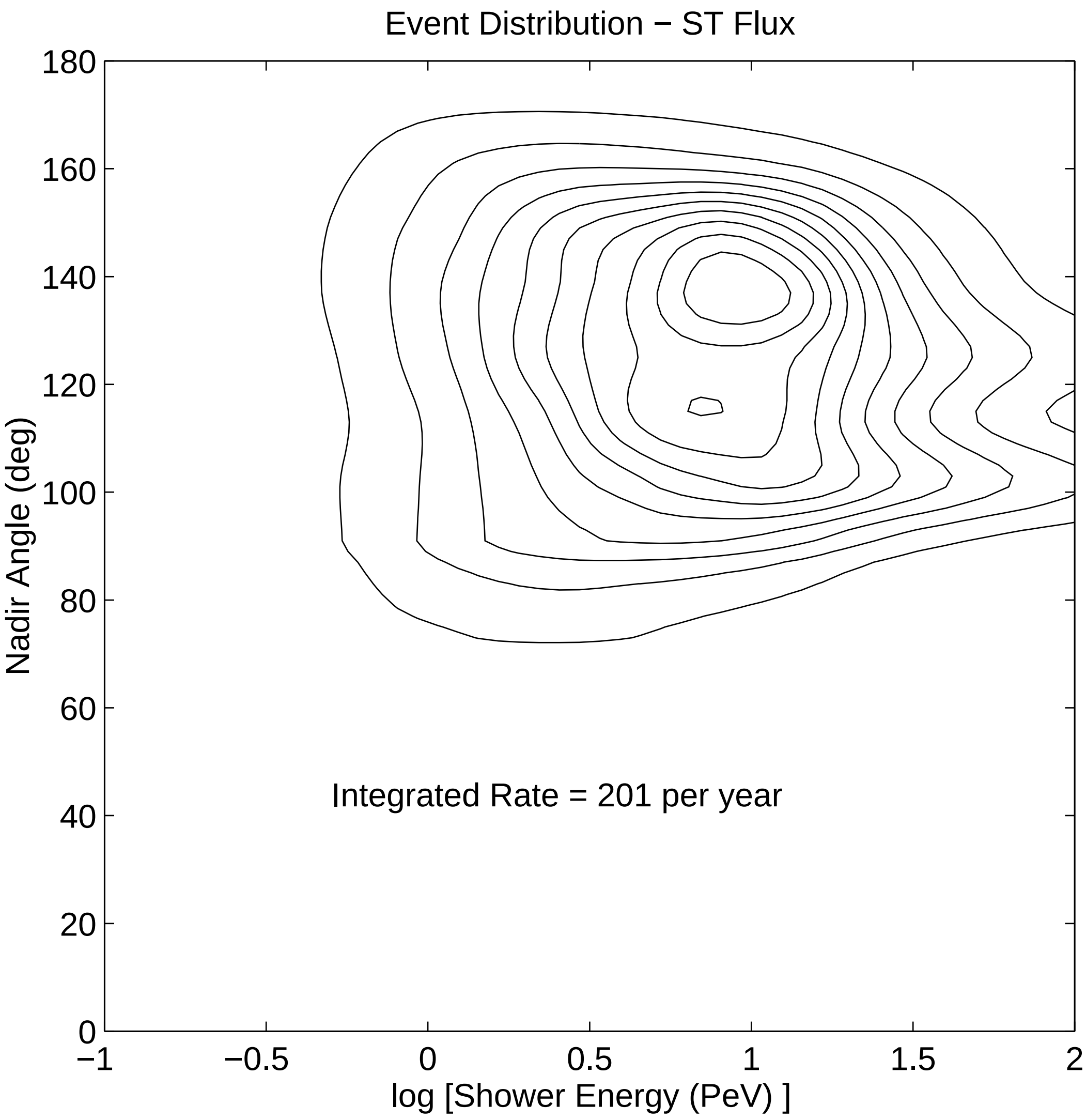

Figure 19: The distribution of AGN neutrino events as detected by a single antenna located 700 meters below the ice surface using the ST flux prediction. We plot 10 equally spaced contours of the quantity $dN/d\cos\theta/d\log E$, with the ratio between the innermost and outermost contours equal to 10. Events are concentrated about an energy of 10 PeV and 140 degree nadir angle. The integrated rate is 201 events per year.

One would always like to have the largest possible number of receivers. However valuable astrophysical information is possible even with a more modest array of sensors. In the case where only hundreds of events are measured, with energy resolution of $\sim 50\%$, say, the distribution peak in energy, $E_{\max}$, would still be estimated with some precision. It can then be used to estimate the neutrino spectrum as follows. The overall rate goes as

$$R(E) \propto V(E) E^{-\gamma} , \tag{14}$$

with $V(E)$ the known detection efficiency (effective volume) and $\gamma$ the unknown integral spectral index. The condition $\frac{dR}{dE} = 0$ applied at the distribution maximum yields

$$\gamma(E_{\max}) = \frac{d\log V}{d\log E} , \tag{15}$$

where we have assumed constant $\gamma$ in the region of the peak.

In Fig.(20) we plot $\gamma(E_{\max})$ using the effective volume shown in Fig.(14) for a nadir angle of 140 degrees, the angle where the distributions of Figs.(18) and (19) peak. Fig.(19) indicates $E_{\max} \approx 8$ PeV which corresponds to a spectrum going like $\gamma = 1$; this is indeed the case for the ST flux at this energy. The SP case shown in Fig.(18) has $E_{\max} \approx 1$ PeV. This corresponds to the spectral index $\gamma = 1.8$ according to Fig.(20). A look at Fig.(2) reveals that 1 PeV is the location of a transition in the SP spectrum from $\gamma = 1$ below 1 PeV to $\gamma = 2.35$ above. The $\gamma = 1$ region 'wants' to have its peak near 10 PeV as in the ST case, and the $\gamma = 2.35$ region 'wants' to peak at about 300 TeV. The net result is an abrupt peak at the transition point with our estimate $\gamma = 1.8$ indicating an intermediate value. The information that the peak in the SP spectrum represents a 'knee' in the spectrum rather than a constant is contained in the shape of the distribution and cannot be determined from the peak location alone.

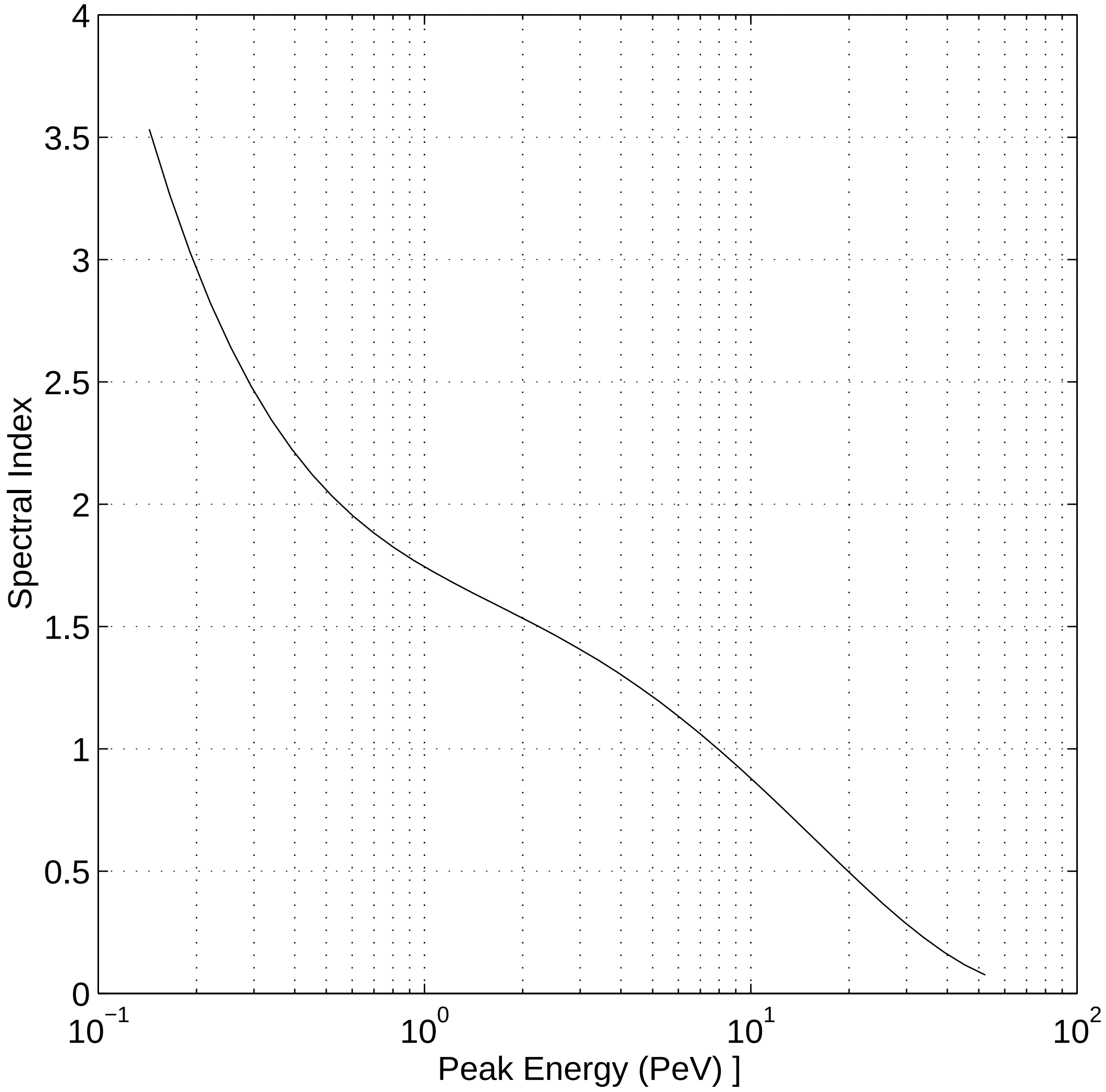

Figure 20: Relation between the peak event distribution energy, $E_{\rm max}$, and the integral spectral index for EM showers, $\gamma(E_{\rm max})$. This spectral index is in turn closely related to the primary neutrino spectrum.

## 12 Summary of Results and Outlook

We have presented a detailed analysis of the problem of radio detection of UHE cosmic ray neutrinos in Antarctic ice. This study, motivated by the exciting prospect of design and construction of a $km^3$ scale detector, is the most thorough to date. We have included every known effect: particle physics in the fundamental interaction cross sections, the angular dependence of attenuation in the Earth, electromagnetic shower evolution, coherent Cherenkov radio emission, polarization of the emitted field, pulse dispersion in the ice and antenna characteristics. Assuming primary UHE fluxes from AGN as calculated by Stecker *et al.* and by Szabo and Protheroe, we calculate very conservatively that there will be at least 200 events recorded per detector element per year. Our results show the tremendous potential for the radio detection technique.

Compared to earlier studies by our group and others, this work has made a significant improvement by considering buried detectors. The main advantage of radio detection is the long range of the radio signal in cold ice, which allows one to detect events efficiently more than a kilometer from the source. In earlier work, it was found that the Earth's presence severely attenuates the interesting PeV neutrinos due to QCD effects that enhance the UHE neutrino-nucleon cross section. Our recent re-analysis of this cross section using new HERA data shows that previous calculations are more than a factor two too low, so the consideration of the Earth's screening is even more important than previously thought. The screening makes receivers which focus strictly on upward-moving neutrinos less effective just in the region where radio has special advantages. By studying the response of buried detectors as a function of deployment depth and incident neutrino angle, we have found an optimum depth at roughly 600 meters where one achieves a substantial increase in rate. The increase in rate comes partly from including detection of sidewise and downward moving primaries and partly from our better treatment of the geometrical factors. The background from atmospheric neutrinos is found to be negligible, as in previous studies of downward looking, surface detectors. A biconical antenna design was chosen for illustration: it has dimensions compatible with phototube deployment and good frequency and angular response for our purpose, as discussed in Sections 6 and 7. A single antenna could detect AGN in either of the two neutrino flux models we use. We have also shown (Section 11) that even a modest data sample with energy resolution on a logarithmic scale allows one to resolve the characteristic spectral index of the source as well. This technique can distinguish clearly between the two flux models considered here, as we have illustrated.

We now turn to the role of muons, discussed in Section 5, and the additional effects of hadronic showers. Recall that muons contributed a small effect from bremsstrahlung, and that their hadronic showers were ignored in the calculations we presented. Our calculations, which are based on the reliably modeled electron-induced showers, establish a firm lower bound on the capabilities of radio detection. It is not difficult to estimate the additional enhancement from including hadronic showers if the assumptions of Provorov and Zheleznykh are used. In the best case that the hadronic shower adds its detectable electromagnetic energy to the electron shower energy linearly, the electron events then approximately double their shower energy. Moreover, adding the muon's hadronic shower (and still treating the muon itself with continuous energy loss) allows one, in effect, to add an electron shower to every muon neutrino induced event. Finally, including neutral currents and the scaling of the effective detection volume with energy, we estimate that each antenna would detect from 4 (Stecker et al. model) to 5 ( Protheroe and Szabo model) times more detected events than those we present in the body of the paper. To use these numbers, the reader can multiply any of the relevant curves by the corresponding factor: for example, the integrated rate for the SP flux from Fig.(18) becomes roughly 2200 events per year, per detector element.

As a last illustrative exercise, we would like to present an estimate for an antenna array - a model for a PeV-scale neutrino telescope. There are many detailed considerations in array design that we do not attempt to address. Consulting Fig.(13), we note that approximately ten detectors arranged a hundred meters or so apart on a vertical string roughly a kilometer long would have non-overlapping volumes and linearly additive rates. This guides our estimate. Ten such strings separated by 1/2 km would seem to be a reasonable array. In a real telescope, of course, there are a number of factors that determine the optimum overlap between receivers. For purposes of discussion, we have simply included a 2:1 redundancy on each detector volume - that is, a detector array with an effective volume 1/2 the value of the sum of the volumes of the individual, widely spaced detectors. We believe this is conservative, because correlations and signal processing can improve the detectable signal-to-noise ratio of a group of receivers to a value well below that of a single receiver, thus increasing the effective detection volume per antenna [35]. We do not include this effect in our estimate; in any case the reader can scale our results by any suitable redundancy factor in a straightforward way. In Fig.(21) we present the event rates calculated with such an array for the SP and ST models, showing both our conservative lower limits (no hadronic showers, few muon neutrino -induced showers) and the optimistic upper limit ( including hadronic showers and neutral current effects). The upper limits are 22,000 and 10,000 for SP and ST, respectively, representing good statistics that might be sufficient for creating true "images" of the PeV neutrino sky!

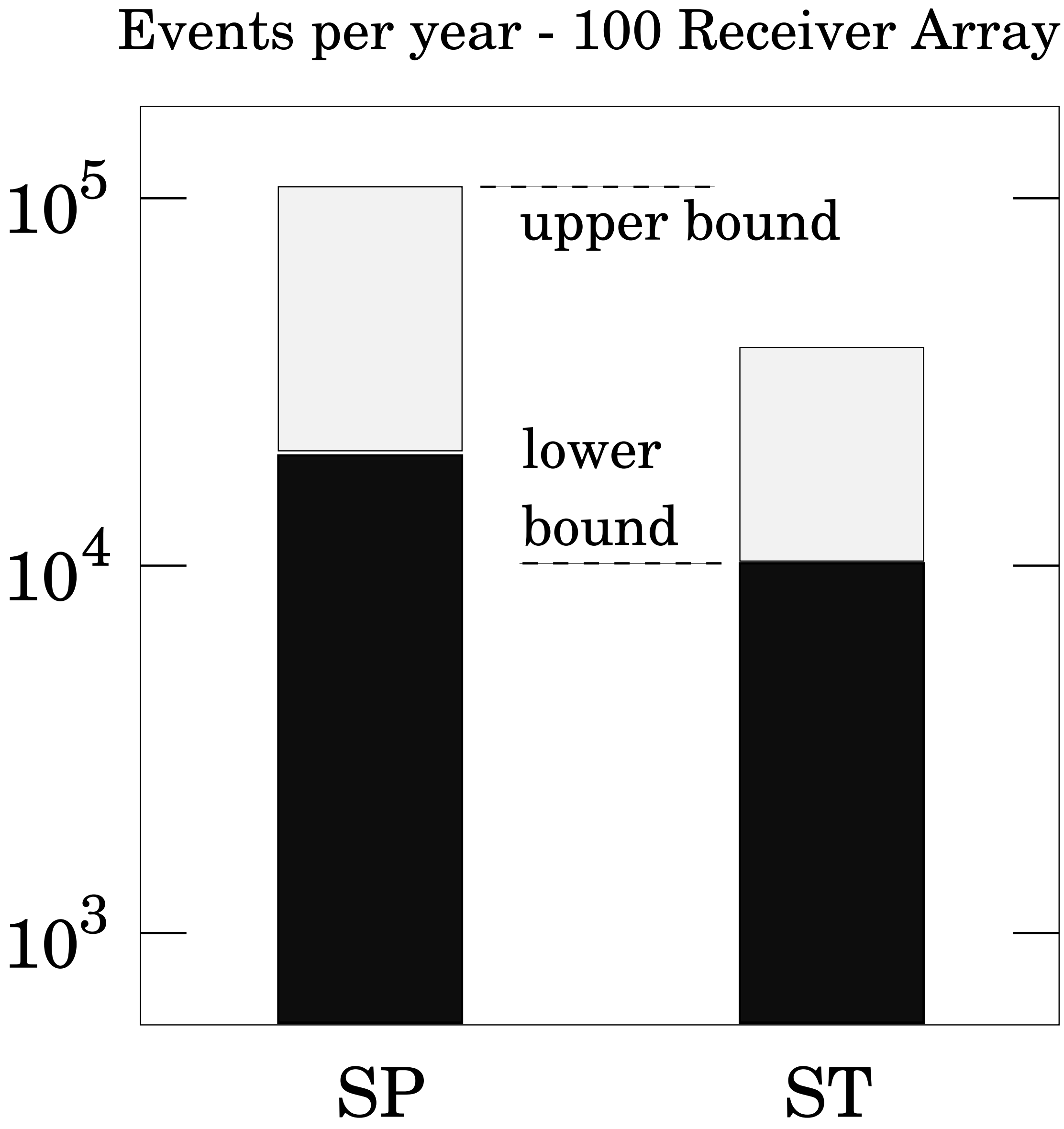

Figure 21: Summary of detected events per year using the SP and ST AGN flux predictions for a 100 receiver array. The volume redundancy we assume is 50%, that is, each observed event appears in two receivers on average. The dark shaded bars represent a lower bound on the observed rate based on pure EM showers induced by charged-current neutrino interactions. The lightly shaded bars show an upper bound on the rate which would occur if the energy in hadronic showers contributes to the signal as efficiently as the pure electromagnetic component. A detailed treatment of hadronic showers should yield a rate between the upper and lower bounds.

The prospects of scientific pay-off by exploiting radio detection of UHE neutrinos are excellent, and this method's complementarity with standard methods is very attractive. We hope that our results will contribute to the push for a pilot experiment to begin the study of UHE neutrinos as soon as possible.

# Acknowledgements

The authors wish to thank E. Zas, F. Halzen, R. Moore, A. Provorov, J. Learned and G. Smoot who each provided valuable assistance at various stages of this project. The work was supported in part under Department of Energy Grant Number DE-FGO2-85-ER 40214 and by the *Kansas Institute for Theoretical and Computational Science* through the K*STAR/NSF program.

# References

[1] T. Gaisser, F. Halzen and T. Stanev,"Particle Astrophysics with High Energy Neutrinos", MAD/PH/847, August 1994. The authors review prospects for 0.1 $km^3$ and 1 $km^3$ detectors briefly.

[2] For a review, see T. Gaisser, *Cosmic Rays and Particle Physics* (Cambridge University Press, 1990).

[3] D. W. McKay and J. P. Ralston, Phys. Lett. **167B**, 103 (1986).

[4] M. H. Reno and C. Quigg, Phys. Rev. **D37**, 657 (1987); see also C. Quigg, M. H. Reno and T. P. Walker, Phys. Rev. Lett. **57**, 774 (1986).

[5] M. H. Reno in *Proceedings of the Workshop on High Energy Neutrino Astrophysics*, Honolulu, Hawaii (23-26 March, 1992) World Scientific (1992), edited by V. J. Stenger, J. G. Learned, S. Pakvasa and X. Tata.

[6] G. M. Frichter, D. W. McKay and J. P. Ralston, Phys. Rev. Lett. **74**, 1508 (1995).

[7] J. G. Learned and S. Pakvasa, Astropart. Phys. **3**, 267 (1995).

[8] G. A. Askaryan, Sov. Phys. JETP **14**, 441 (1962).

[9] H. R. Allan in *Progress in Elementary Particle and Cosmic Ray Physics*, edited by J. G. Wilson and S. A. Wouthuysen, North-Holland, Amsterdam (1971).

[10] N. Mandolisi, G. Morigi and G. Palumbo, J. Phys. A **9**, 815 (1976); H. Allen *et al.*, *Proceedings of the 14th International Conference on Cosmic Rays*, Munich Vol. 8, 3077-81 (1975).

[11] M. Markov and I. Zheleznykh, Nucl. Instrum. Meth. Phys. Res. **A248**, 242 (1986).

[12] I. Zheleznykh, in *Proceedings of Neutrino '88*, Boston-Medford 1988, p. 528, edited by J. Schneps, T. Kafka and W. A. Mann (World Scientific 1989); I. Zheleznykh, in *Proceedings of the 21st International Cosmic Ray Conference*, Adelaide 1989, Vol. 6, p. 528-533.

[13] J. P. Ralston and D. W. McKay, *Astrophysics in Antarctica*, AIP Conference Proceedings **198**, 24 (1990); J. P. Ralston and D. W. McKay, *Arkansas Gamma-Ray and Neutrino Workshop - 1989*, Little Rock 1989, Nucl. Phys. B (Proc. Suppl.) **14A**, 356 (1990).

[14] E. Zas, F. Halzen and T. Stanev, Phys. Rev. **D 45**, 362 (1992).

[15] A. L. Provorov and I. Zheleznykh, *Proceedings of the 23rd International Cosmic Ray Conference*, Calgary 1993; A. L. Provorov and I. Zheleznykh, "Radiowave Method of High Energy Neutrino Detection:

Calculation of Expected Event Rate", to appear in Astropart. Phys. October 1995.

[16] See Proceedings of *Snowmass 94* Meeting on Neutrino Astrophysics, July 1994.

[17] V. V. Bogorodsky and V. P. Gavrilo, *Ice: Physical Properties*, Modern Mothods of Glaciology (Lenningrad, 1980).

[18] T. Stanev and H. Vankov, Phys. Rev. D **40**, 1472 (1089).

[19] F. W. Stecker, C. Done, M. H. Salamon and P. Sommers, Phys. Rev. Lett. **66**, 2697; E (1992): *ibid* **69**, 2738.

[20] A. P. Szabo and R. J. Protheroe, Astropart. Phys. **2**, 375 (1994); R. J. Protheroe and A. P. Szabo, Phys. Rev. Lett. **69**, 2285 (1992).

[21] F. W. Stecker and M. H. Salamon, "High Energy Neutrinos from Quasars", NASA/Goddard preprint ASTRO-PH-9501064 (1995), to appear in Space Science Reviews.

[22] For AGN neutrino detection studies, see, e.g. J. Learned and V. Stenger; F. Halzen and E. Zas; M. C. Goodman *et al.*; A. Okada in *Proceedings of the Workshop on High Energy Neutrino Astrophysics*, Honolulu, Hawaii (23-26 March, 1992) World Scientific (1992), edited by V. J. Stenger, J. G. Learned, S. Pakvasa and X. Tata.

[23] P. Lipari, Astropart. Phys. **1**, 195 (1993). We have used the $90^o$ flux and extrapolated slightly beyond the $10^2 - 10^3$ TeV results of this paper.

[24] H1 Collaboration, Nucl. Phys. **B407**, 515 (1993).

[25] ZEUS Collaboration, Phys. Lett. **316B**, 412 (1993); ZEUS Collaboration, preprint DESY 94-143 (1994).

[26] H1 Collaboration, Phys. Lett. **324B**, 241 (1994).

[27] V. S. Berezinsky, A. Z. Gazizov, A. Z. Rozental and G. T. Zatsepin, Sov. J. Nucl. Phys. **43**, 406 (1986).

[28] A. M. Dziewonski and D. L. Anderson, Phys. Earth Planet. Inter. **25**, 297 (1981).

[29] R. K. Adair and H. Kasha, in "Muon Physics: I Electromagnetic Interactions", edited by V. W. Hughes and C. S. Wu, Academic Press 1977.

[30] A. N. Salamantin *et al.*, in *Antarctic Committee Reports*, Vol. 24, 94 (Moscow, 1985).

[31] C. W. Harrison and C. S. Williams, IEEE Trans. Anten. Prop., **AP-13**, 236 (1965); M. Kanda, IEEE Trans. Elec. Compat., **EMC-24**, 245 (1982); C. H. Papas and R. King, IRE Proc., vol.**39**, 1269 (1949). We thank Dick Moore for suggesting we consider biconical antennas.

[32] C. T. Tai and S. A. Long in: *Antenna Engineering Handbook*, McGraw-Hill, New York (1984), edited by R. C. Johnson and H. Jasik; G. H. Brown and O. M. Woodward, Jr., RCA Review, vol.**13**, no.4, 425 (1952).

[33] I. N. Boldyrev, G. A. Gusev, M. A. Markov, A. L. Provorov and I. M. Zeleznykh, *Proc. XXth International Cosmic Ray Conference*, Vol. 6, 472 (Moscow 1987).

[34] J. Minkoff, *Signals, Noise, and Active Sensors*, John Wiley & Sons, Inc., New York (1992).

[35] Monte Carlo studies for radio detector design optimization are being carried out by A. Bean, D. Besson, S. Kotov, I. Kravchenko and S. Seunarine at the University of Kansas. Alice Bean and Dave Besson, private communication.